\documentclass[aps,prb,showpacs,reprint,superscriptaddress,floatfix]{revtex4-1}

\usepackage{graphicx}
\usepackage{color}
\usepackage{amsmath}
\usepackage{bbm}
\usepackage{amssymb}

\usepackage{dsfont}

\usepackage[utf8]{inputenc}	% kodowanie UTF-8
\usepackage[T1]{fontenc}

\input epsf

\begin{document}

%Authors
\title{Electric dipole spin resonance in single and two electron quantum dot defined in two-dimensional electron gas at the SrTiO$_3$/LaAlO$_3$ interface.}

\author{B. Szafran}
\email{bszafran@agh.edu.pl}
\affiliation{AGH University of Krakow, Faculty of Physics and Applied Computer Science, Al. Mickiewicza 30, 30-059 Krakow, Poland}

\author{M. Zegrodnik}
\email{michal.zegrodnik@agh.edu.pl}
\affiliation{AGH University of Krakow, Academic Centre for Materials and Nanotechnology, Al. Mickiewicza 30, 30-059 Krakow, Poland}

\author{M. P. Nowak}
\email{mpnowak@agh.edu.pl}
\affiliation{AGH University of Krakow, Academic Centre for Materials and Nanotechnology, Al. Mickiewicza 30, 30-059 Krakow, Poland}

\author{R. Citro}
\email{rocitro@unisa.it}
\affiliation{Department of Physics ''E.R. Caianiello'' University of Salerno and CNR-SPIN, Via Giovanni Paolo II, 132, Fisciano (Sa), Italy}

\author{P. W\'ojcik}
\email{pawelwojcik@agh.edu.pl}
\affiliation{AGH University of Krakow, Faculty of Physics and Applied Computer Science, Al. Mickiewicza 30, 30-059 Krakow, Poland}

\date{February 2023}

\begin{abstract}
We investigate the energy spectrum of a single and two electron quantum dot (QD) embedded in two dimensional electron gas at the interface between SrTiO$_3$ and LaAlO$_3$, in the presence of the external magnetic field. For this purpose the three band model of $3d$-electrons defined on the square lattice of Ti ions was utilized. We demonstrate that, for the weak parabolic confinement potential, the low energy spectrum is sufficiently well described by the effective Hamiltonian reduced to the one $d_{xy}$ orbital with the spin-orbit interaction originating from the coupling to the $d_{xz}$, $d_{yz}$ bands. This is not the case for stronger confinement where contribution of the states related to the $d_{xz/yz}$ orbital is relevant. Based on the time depended calculations we discuss in details the manipulation of the electron spin in QD by external AC voltages, in the context of the electric dipole spin resonance.
The allowed and forbidden transitions are discussed in details with respect to the parity selection rule.  Our calculations show that for a single electron QD  the spin-flip in the ground-state has a character of a Rabi resonance while for two electrons the singlet-triplet transition is forbidden by the parity symmetry. For the two electrons QD, we demonstrate that the spin-flip transition can still be accomplished via a second-order, two-photon process that has a two-state Rabi character for low AC field amplitude. The violation of the parity symmetry on the spin-flip transitions is also analyzed.
\end{abstract}

\maketitle

\section{Introduction}
The spin dynamics of electrons confined in quantum dots (QDs) has attracted an increasing interest in recent years as a fundamental aspect for constructing spin qubits for future quantum information processing~\cite{Burkard2023,Nadj-Perge2012}. The key element in utilizing spin qubits in quantum technology is an effective control of their states through the coherent spin manipulation. In the case of a single quantum dot, this can be realized through the electron spin resonance-Rabi oscillations induced by an external microwave radiation, which drives resonant transitions between the Zeeman-split energy levels in a magnetic field\cite{Koppens2005,Koppens2006, Koppens2008}. Alternatively, two-electron spin states in QDs have been demonstrated to be effectively tunable by exchange coupling\cite{Petta2005}. 

A significant breakthrough in this field has been achieved in electrostatically defined quantum dots \cite{Nadj-Perge2012_2, Schroer2011}, where the microwave field has been successfully replaced by AC gate voltages\cite{Nowack2007,Nadj-Perge2010}. Periodic changes in the potential induced by the AC field affect the spin of the confined electron through the momentum-dependent spin-orbit (SO) interaction\cite{Bychkov1984,Rashba2003}. This technique, called electric-dipole spin resonance (EDSR)\cite{Golovach2006, Nowak2012, Sherman1, Sherman2}, has been reported in two-electron double-quantum-dot systems using Pauli blockade of the current flow, which occurs when the dots are occupied by electrons with parallel spins\cite{Nowack2007,Nadj-Perge2010}. 

The number of operations which can be performed on a spin-qubit is a result of the switching time (mainly determined by the SO coupling strength) and the spin decoherence. The latter is strongly influenced by the host material, which sets the strength of hyperfine interaction with nuclear spins as well as the coupling of the electron spin with the lattice vibrations and charge fluctuations via the SO interaction. Thus, the SO coupling, on the one hand, is responsible for coherent spin manipulations, but on the other hand it is also the source of detrimental spin decoherence. 

The two-dimensional electron gas (2DEG) formed at the interface between SrTiO$_3$ (STO) and other insulating transition metal oxides, such as LaAlO$_3$ (LAO) \cite{levy1} is considered a promising material platform for the QD-based spin qubits fabrication. The increasing interest in this platform is mainly due to the fact that the considered 2DEG  interface exhibits a unique combination of characteristics, including high mobility\cite{Ohtomo2004}, large spin-orbit coupling\cite{Caviglia2010, Yin2020}, gate-tunable superconductivity\cite{Maniv2015,Monteiro2019,Joshua2012,Biscaras2012}, magnetic ordering\cite{Pavlenko2012}, and ferroelectricity\cite{Noel2020}. Its susceptibility to electrostatic gating, comparable to semiconductor materials, has brought significant advances in oxide-2DEG nanotechnology\cite{Jespersen2020,Thierschmann2020,Guenevere2017,Jouan2020}. A recent study of single QDs based on LAO/STO reveals the Coulomb blockade diamonds characteristic for well-defined electrostatic confinement~\cite{Jespersen2020}. The electron spin in the STO-based QDs can be controlled by a large SO coupling, switched on and off by moderate gate voltages, which is hardly achievable with semiconducting platforms\cite{Caviglia2010}. Note, moreover, that the STO-based 2DEG is expected to have another significant advantage over semiconductors. It is characterized by a smaller impact of direct and indirect sources of decoherence, as the hyperfine interaction with the nuclear spin bath should be suppressed by 3d-character of electrons - their wave functions have node at the nuclei positions\cite{Loss2002}. All these properties, combined with the capability to interconvert charge and spin currents through Edelstein and spin Hall effects\cite{Lesne2016,Vaz2019,Trama2022, Trama2022_2}, exhibiting some of the highest efficiencies among solid-state materials, make LAO/STO-based quantum dots a promising platform for a development of fast spin qubits with inherent scalability of 2D systems. However, so far, the electronic structure and spin dynamics in the STO-based QDs have not been systematically explored. Note, that the structural lateral confinement of the 2DEG
at the SrTiO$_3$/LaAlO$_3$ interface using nanolitography has also been applied \cite{Maniv}
for formation of QDs.

In this paper, based on the three band model we investigate an electronic spectrum of a single and two electrons STO-based QD. We derive the simplified one band Hamiltonian for the $d_{xy}$ orbital with inclusion of the SO interaction and demonstrate that it can sufficiently well describe the low energy range of the spectrum for the low parabolic confinement. Based on the time dependent scheme we simulate the electric dipole spin resonance with the electron spin in QD controlled by the AC electric field. The calculated transitions between states are discussed with respect to the direct (single photon) and second order (two-photon) processes determined by the parity symmetry of states. The organization of the manuscript is the following: in Sec. II we present a theoretical model in the $k$-vector and real space as well as the simplified Hamiltonian for the $d_{xy}$ band, Sec. II contains the analysis of electronic spectrum of a single and two electron QD as well as results of time dependent simulations (EDSR), finally summary and conclusions are included in Sec. IV. 

\section{Theoretical model}

\subsection{Single-electron Hamiltonian for (001)-oriented LAO/STO interface.}
At the (001)-oriented LAO/STO, the conduction band is formed by the Ti $t_{2g}$ orbitals ($d_{dxy}$, $d_{yz}$, $d_{xz}$) coupled through $2p$ states of oxygen on the square lattice. At the interface, where a narrow quantum well is created\cite{Popovic2008,Pentcheva2006,Pavlenko2012} as a result of the polarization discontinuity, the degeneracy of the $t_{2g}$ bands is lifted, resulting in 2D discrete states with the band $d_{xy}$ being lower in energy with respect to the bands $d_{yz}$, $d_{xz}$. In wave vector space, 2DEG at the (001) LAO/STO interface can be described by the Hamiltonian\cite{Diez2015}
\begin{equation}
    \hat{H}_{\mathbf{k}}=\sum _{\mathbf{k}} \hat{C}^\dagger_{\mathbf{k}} ( \hat{H}_{0}+\hat{H}_{RSO}+\hat{H}_{SO}+\hat{H}_B ) \hat{C}_{\mathbf{k}},
    \label{eq:Hamiltonian_k_space}
\end{equation}
where $\hat{C}_{\mathbf{k}}=(\hat{c}_{\mathbf{k},xy}^{\uparrow}, \hat{c}_{\mathbf{k},xy}^{\downarrow}, \hat{c}_{\mathbf{k},xz}^{\uparrow}, \hat{c}_{\mathbf{k},xz}^{\downarrow}, \hat{c}_{\mathbf{k},yz}^{\uparrow}, \hat{c}_{\mathbf{k},yz}^{\downarrow})^{T}$ corresponds to the vector of anihilation operators for electrons with spin $\sigma=\uparrow,\downarrow$ on the orbital $d_{xy},d_{xz},d_{yz}$, in the state $\mathbf{k}$.  In Eq.~(\ref{eq:Hamiltonian_k_space}), $\hat{H}_0$ describes the kinetic energy
and is given by
\begin{equation}
\hat{H}_{0}=
\left(
\begin{array}{ccc}
 \epsilon^{xy}_{\mathbf{k}} & 0 & 0\\
 0 & \epsilon^{xz}_{\mathbf{k}} &  \epsilon^h_{\mathbf{k}} \\
 0 & \epsilon^h_{\mathbf{k}}  &  \epsilon^{yz}_{\mathbf{k}}
\end{array} \right) \otimes \hat {\sigma} _0\;,
\end{equation}
with dispersion relations 
\begin{equation}
\begin{split}
    \epsilon^{xy}_{\mathbf{k}}&=4t_l-2t_l\cos{k_x}-2t_l\cos{k_y}-\Delta_E,\\
    \epsilon^{xz}_{\mathbf{k}}&=2t_l+2t_h-2t_l\cos{k_x}-2t_h\cos{k_y},\\
    \epsilon^{yz}_{\mathbf{k}}&=2t_l+2t_h-2t_h\cos{k_x}-2t_l\cos{k_y},
\end{split}
\label{eq:H0}
\end{equation}
and the hybridization term defined by 
\begin{equation}
\epsilon^h_{\mathbf{k}}=2t_d\sin{k_x}\sin{k_y}.
\end{equation}
In Eq. (3) and (4) $t_l$, $t_h$ are the hopping parameters (energies)  for the light and heavy mass and $t_d$ is the energy which determines the coupling between the $d_{xz}/d_{yz}$ band.
Due to the lack of the inversion symmetry occurring in a natural way at interfaces, LAO/STO-based 2DEG exhibits the SO coupling, in this case, consisting of two components: the atomic and Rashba part. The former appears as an effect of the atomic $\mathbf{L} \cdot \mathbf{S}$ interaction and can be expressed in the form~\cite{Khalsa2013}
\begin{equation}
\hat{H}_{SO}= \frac{\Delta_{SO}}{3}
\left(
\begin{array}{ccc}
0 & i \sigma _x & -i \sigma _y\\
-i \sigma _x & 0 & i \sigma _z \\
i \sigma _y & -i \sigma _z & 0
\end{array} \right) \;,
\label{eq:hso}
\end{equation}
where $\Delta _{SO}$ determines the strength of the atomic spin-orbit energy and $\sigma _{x},\sigma_{y},\sigma_{z}$ are the Pauli matrices. \\
The Rashba-like spin-orbit term $\hat{H}_{RSO}$, occurring as a result of the mirror symmetry breaking, is induced by the out-of-plane offset of atom positions at the interface and is given by
\begin{equation}
\hat{H}_{RSO}= \Delta_{RSO}
\left(
\begin{array}{ccc}
0 & i \sin{k_y} & i \sin{k_x}\\
-i \sin{k_y} & 0 & 0 \\
-i \sin{k_x} & 0 & 0
\end{array} \right) \otimes \hat {\sigma} _0\;,
\label{eq:rso}
\end{equation}
where $\Delta _{RSO}$ determines the energy of the Rashba SO coupling. \\

Finally, the coupling of the external magnetic field to the spin and orbital momentum of electrons is taken into account by the Hamiltonian
\begin{equation}
\hat{H}_B=\mu_B(\mathbf{L}\otimes \sigma_0+g\mathds{1}_{3\times 3} \otimes \mathbf{S})\cdot \mathbf{B}/\hbar,
\label{eq:Hb}
\end{equation}  
where $\mu_B$ is the Bohr magneton, $g$ is the Land\'e factor, $\mathbf{S}=\hbar \pmb{\sigma}/2$ with $\pmb{\sigma}=(\sigma_x,\sigma_y,\sigma_z)$ and $\mathbf{L}=(L_x,L_y,L_z)$ with
\begin{equation}
\begin{split}
 L_x&= \left ( 
 \begin{array}{ccc}
  0 & i & 0 \\
  -i & 0 & 0 \\
  0 & 0 & 0 
 \end{array}
 \right ), 
 L_y= \left ( 
 \begin{array}{ccc}
  0 & 0 & -i \\
  0 & 0 & 0 \\
  i & 0 & 0 
 \end{array}
 \right ), 
 L_z= \left ( 
 \begin{array}{ccc}
  0 & 0 & 0 \\
  0 & 0 & i \\
  0 & -i & 0 
 \end{array}
 \right ).
 \end{split}
\end{equation}

\begin{figure}[!htp]
    \centering
    \includegraphics[width = 0.5 \textwidth]{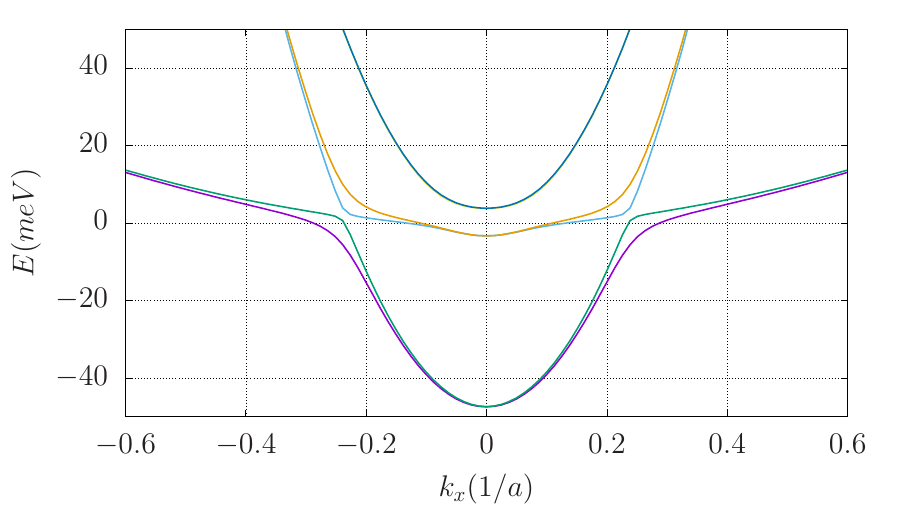}
    \caption{Dispersion relation $E(k_x,k_y=0)$ for the (001)-oriented LAO/STO heterostructures.}
    \label{fig1}
\end{figure}

In our calculations, we assume the tight-binding parameters $t_l=875\;$meV, $t_h=40\;$meV, $t_d=40\;$meV, $\Delta_E=47\;$meV taken from Ref.~\onlinecite{Maniv2015}, the Land\'e factor~\cite{Ruhman2014} $g=3$ and the SO coupling parameters $\Delta_{SO}=10$~meV, $\Delta_{RSO}=20$~meV corresponding to that measured experimentally\cite{Caviglia2010,Yin2020}.
The dispersion relation $E(k)$ for the chosen parameters is presented in Fig.~\ref{fig1}.\\

For confined nanostructures, such as QDs it is necessary to express Hamiltonian (\ref{eq:Hamiltonian_k_space}) in the real space. On the square lattice, with the position indexed by $(\mu, \nu)$, the single-electron Hamiltonian (\ref{eq:Hamiltonian_k_space}) takes the form 
\begin{eqnarray}
    \hat{H}&=&\sum _{\mu,\nu} \hat{C}^\dagger_{\mu, \nu} ( \hat{H}^{0}+\hat{H}_{SO}+\hat{H}_B ) \hat{C}_{\mu, \nu}+ 
    \label{eq:Hamiltonian_real} \\
    &&\sum _{\mu,\nu} \hat{C}^\dagger_{\mu+1, \nu} \hat{H}^{x} \hat{C}_{\mu, \nu}  + \sum _{\mu,\nu} \hat{C}^\dagger_{\mu, \nu+1} \hat{H}^{y} \hat{C}_{\mu, \nu} + \nonumber \\ 
    &+& \sum _{\mu,\nu} \hat{C}^\dagger_{\mu+1, \nu-1} \hat{H}_{mix} \hat{C}_{\mu, \nu}-
    \sum _{\mu,\nu} \hat{C}^\dagger_{\mu+1, \nu+1} \hat{H}_{mix} \hat{C}_{\mu, \nu} +h.c., \nonumber
\end{eqnarray}
where $\hat{C}_{\mu, \nu}=(\hat{c}_{\mu, \nu,xy}^{\uparrow}, \hat{c}_{\mu, \nu, xy}^{\downarrow}, \hat{c}_{\mu, \nu, xz}^{\uparrow}, \hat{c}_{\mu, \nu, xz}^{\downarrow}, \hat{c}_{\mu, \nu, yz}^{\uparrow}, \hat{c}_{\mu, \nu, yz}^{\downarrow})^{T}$ corresponds to the vector of anihilation operators of electron with spin $\sigma=\uparrow,\downarrow$ on the orbital $d_{xy}, d_{xz}, d_{yz}$ and the position $( \mu,\nu )$, $\hat{H}^{0}$ defines the on-site energy related to the kinetic term and the confinement potential $V(\mathbf{r})$ 
\begin{eqnarray}
\hat{H}^{0}&=&
\left(
\begin{array}{ccc}
 4t_l-\Delta _E & 0 & 0\\
 0 & 2t_l+2t_h & 0 \\
 0 & 0  &  2t_l+2t_h
\end{array} \right) \otimes \hat {\sigma} _0 \nonumber \\
&+&
\left(
\begin{array}{ccc}
 V_{\mu,\nu} & 0 & 0\\
 0 & V_{\mu,\nu} & 0 \\
 0 & 0  &  V_{\mu,\nu}
\end{array} \right) \otimes \hat {\sigma} _0 ,
\label{eq:Hamiltonian_real_H0}
\end{eqnarray}
while 
\begin{equation}
\hat{H}^{x}=
\left(
\begin{array}{ccc}
 -t_l & 0 & 0\\
 0 & -t_l & 0 \\
 0 & 0  &  -t_h
\end{array} \right) \otimes \hat {\sigma} _0
+
\frac{\Delta_{RSO}}{2}
\left(
\begin{array}{ccc}
 0 & 0 & -1 \\
 0 & 0 & 0 \\
 1 & 0  &  0
\end{array} \right) \otimes \hat {\sigma} _0\;,
\label{eq:real_space_Hx}
\end{equation}
\begin{equation}
\hat{H}^{y}=
\left(
\begin{array}{ccc}
 -t_l & 0 & 0\\
 0 & -t_h & 0 \\
 0 & 0  &  -t_l
\end{array} \right) \otimes \hat {\sigma} _0
+
\frac{\Delta_{RSO}}{2}
\left(
\begin{array}{ccc}
 0 & -1 & 0 \\
 1 & 0 & 0 \\
 0 & 0  &  0
\end{array} \right) \otimes \hat {\sigma} _0,
\label{eq:real_space_Hy}
\end{equation}

\begin{equation}
\hat{H}_{mix}= \frac{t_d}{2}
\left(
\begin{array}{ccc}
 0 & 0 & 0 \\
 0 & 0 & 1 \\
 0 & 1  &  0
\end{array} \right) \otimes \hat {\sigma} _0 ,
\label{eq:real_space_hybrid}
\end{equation}
determines the energy of hopping to the nearest neighbours related to the kinetic energy and Rashba SO coupling (\ref{eq:real_space_Hx}, \ref{eq:real_space_Hy}) as well as the hybridization (\ref{eq:real_space_hybrid}). In Eq.~(\ref{eq:Hamiltonian_real}), $\hat{H}_{SO}$ and $\hat{H}_B$ have the same form as in the wave vector space formulation and are given by Eqs.~(\ref{eq:hso}) and (\ref{eq:Hb}).

\subsection{Simplified Hamiltonian for the $d_{xy}$ band}
The position the $d_{xy}$ band on the energy scale lowered by $\Delta _E$ relative to the $d_{yz} / d_{xz}$ bands, together with the low effective mass, may lead to the dominant role of $d_{xy}$ states when the confinement of QD is weak. In such a case, one can anticipate that the influence of the higher-lying bands, $d_{yz}$ and $d_{xz}$, is negligible. It is essential to note that reducing the Hamiltonian (\ref{eq:Hamiltonian_k_space}) solely to the $d_{xy}$ part is not sufficient, as it overlooks the SO coupling arising from the interaction with the $d_{xz/yz}$ bands via the terms $H_{RSO}$ and $H_{SO}$. For this reason, more sophisticated methods are required to reduce the full Hamiltonian to the effective one describing the $d_{xy}$ band. To derive it, let us express Hamiltonian (\ref{eq:Hamiltonian_k_space}) in the following form
\begin{equation}
\hat{H}=
\left(
\begin{array}{cc}
 \hat{H}_{xy} & \hat{H}_{c}\\
 \hat{H}_c^\dagger & \hat{H}_{xz/yz}
\end{array} \right)-\Delta_E \mathds{1}_{6\times 6},
\label{eq:Hamiltonian_to_reduce}
\end{equation}
where
\begin{equation}
\hat{H}_{xy}=
\left(
\begin{array}{cc}
\epsilon^{xy}_{\mathbf{k}} + \Delta_E & 0\\
 0 & \epsilon^{xy}_{\mathbf{k}}  + \Delta_E
\end{array} \right)+\frac{1}{2}g\mu_B \mathbf{B}\cdot \pmb{\sigma},
\end{equation}
\begin{equation}
\hat{H}_{xz/yz}=
\left(
\begin{array}{cccc}
\epsilon^{xz}_{\mathbf{k}}+\Delta_E & 0 & 0 & 0 \\
 0 & \epsilon^{xz}_{\mathbf{k}}+\Delta_E & 0 & 0 \\
 0 & 0 & \epsilon^{yz}_{\mathbf{k}}+\Delta_E & 0 \\ 
 0 & 0 & 0 & \epsilon^{yz}_{\mathbf{k}}+\Delta_E \\
\end{array} \right),
\end{equation}
while the coupling between the states $d_{xy}$ and $d_{xz}/d_{yz}$ is given by
\begin{eqnarray}
\hat{H}_{c}&=&
\frac{\Delta_{SO}}{3}
\left(
\begin{array}{cccc}
0 & i & 0 & -1 \\
i & 0  & 1 & 0   
\end{array} \right)  \\
&+&i\Delta_{RSO}
\left(
\begin{array}{cccc}
\sin k_y & 0 & \sin k_x & 0 \\
0 & \sin k_y & 0 & \sin k_x   
\end{array} \right). \nonumber
\end{eqnarray}
Note that in $\hat{H}_{xz/yz}$ we neglect the coupling of the bands $d_{xz} / d_{yz}$ to the magnetic field and their hybridization assuming that the kinetic and SO energy constitute the major contribution to the energy. 

Using the standard folding-down transformation, we can reduce the $6\times 6$  model (\ref{eq:Hamiltonian_to_reduce}) into the effective $2\times 2$ Hamiltonian for the $d_{xy}$ electrons
\begin{equation}
    \hat{H}^{eff}_{xy} = \hat{H}_{xy} + \hat{H}_{c}(\hat{H}_{xz/yz} - E)^{-1}\hat{H}_{c}^\dagger.
    \label{Heff}
\end{equation}
If we assume that $\Delta_E$ is much larger than the kinetic energy in the $d_{xz}/d_{yz}$ band, we can expand $(\hat{H}_{xz/yz} - E)^{-1}$ from Eq.~(\ref{Heff}) in the Taylor series 
\begin{equation}
(\hat{H}_{xz/yz} - E)^{-1} = \frac{1}{\epsilon^{xz/yz}_{\mathbf{k}}+\Delta_E-E} \mathds{1}_{4\times4} \approx \frac{1}{\Delta_E}\mathds{1}_{4\times4}.
\end{equation}
Then, we the effective Hamiltonian is reduced to the following form
\begin{eqnarray}
\hat{H}^{eff}_{xy}&=&\left ( \epsilon^{xz}_{\mathbf{k}} + \frac{2\Delta_{SO}\gamma}{3(1-\gamma)}\right ) \mathbf{1}_{2\times2}+\frac{1}{2}g\mu_B \mathbf{B}\cdot \pmb{\sigma} \nonumber \\ 
&+&\alpha (\sigma_y \sin k_x - \sigma _x \sin k_y)
\label{eq:Hamiltonian_k_xy}
\end{eqnarray}
where $\gamma=\Delta_{SO} / 3\Delta _E$ and $\alpha=\Delta_{SO}\Delta_{RSO}/3\Delta_E$. The last term in Eq.~(\ref{eq:Hamiltonian_k_xy}) is related to the SO coupling of the Rashba type \cite{Rashba,Rashba2003} similar to that observed in semiconductor 2DEGs. It clearly demonstrates that the SO coupling for the $d_{xy}$ band results from the relative interplay between the atomic and Rashba coupling through the bands $d_{xz} / d_{yz}$.

The Hamiltonian (\ref{eq:Hamiltonian_k_xy}) in the real space takes the form
\begin{eqnarray}
    \hat{H}_{xy}^{eff}&=&\sum _{\mu,\nu} \hat{C}^\dagger_{\mu, \nu} \left [ \left ( 4t_l -  \frac{2 \Delta_{SO}\gamma}{3(1-\gamma)} \right ) \sigma_0 + \right. \nonumber \\ &+& \left. \frac{1}{2}g \mu_B \mathbf{B} \cdot \pmb{\sigma} +V_{\mu,\nu} \right ] \hat{C}_{\mu, \nu} \nonumber \\
    &+&\sum _{\mu,\nu} \hat{C}^\dagger_{\mu+1, \nu} \left ( -t_l \sigma_0 - i \alpha \sigma_y \right ) \hat{C}_{\mu, \nu}  \nonumber \\
    &+& \sum _{\mu,\nu} \hat{C}^\dagger_{\mu, \nu+1} \left ( -t_l \sigma_0 + i \alpha \sigma_x \right ) \hat{C}_{\mu, \nu} + h.c
    \label{eq:Hamiltonian_real_xy}
\end{eqnarray}
which, as we will show later, can be successfully used to describe the electronic structures of QDs with a weak confinement. In the above expression, $V_{\mu,\nu}$ corresponds to the confining potential of QD.

\begin{figure*}
\begin{tabular}{cccc}
\includegraphics[width=0.4\columnwidth]{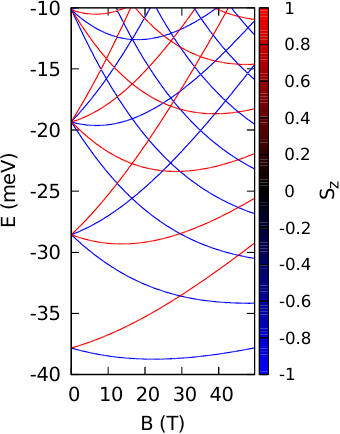} \put(-75,35){(a)}&\includegraphics[width=0.42\columnwidth]{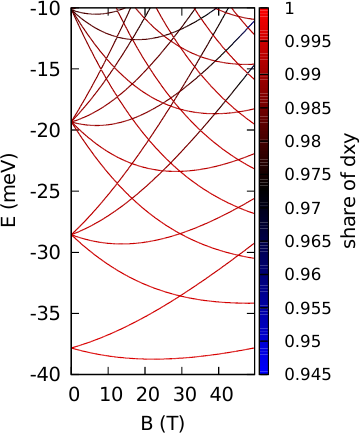}\put(-80,35){(b)}&\includegraphics[width=0.4\columnwidth]{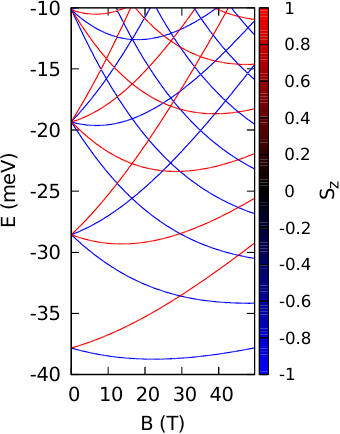}\put(-75,35){(c)}
&\includegraphics[width=0.4\columnwidth]{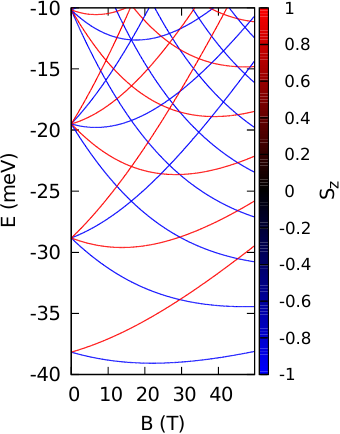}\put(-75,35){(d)}\\
\includegraphics[width=0.4\columnwidth]{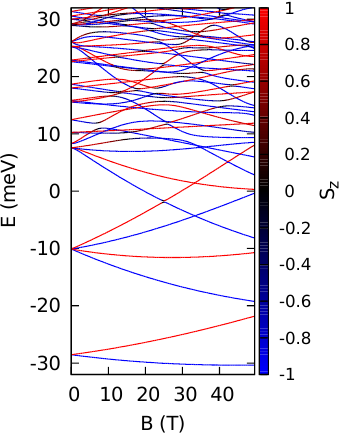}\put(-75,35){(e)}&\includegraphics[width=0.4\columnwidth]{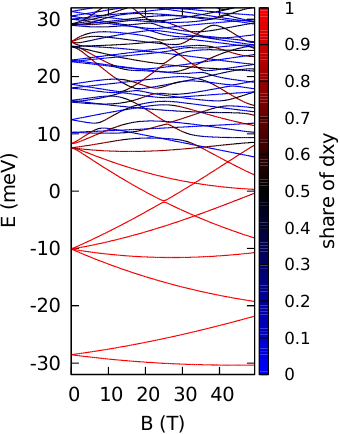}\put(-75,35){(f)}&\includegraphics[width=0.4\columnwidth]{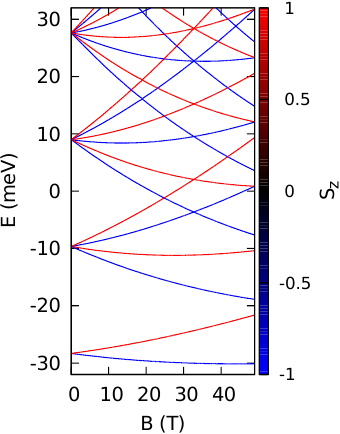}\put(-75,35){(g)}
&\includegraphics[width=0.4\columnwidth]{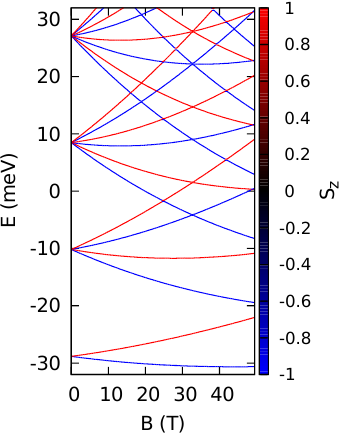}\put(-75,35){(h)}
\\
\includegraphics[width=0.4\columnwidth]{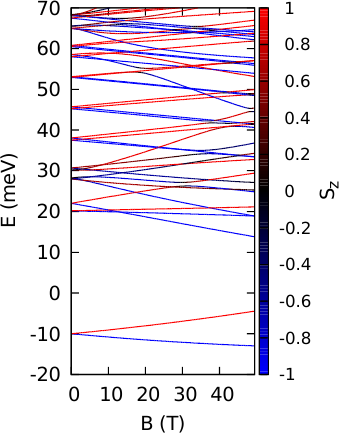}\put(-75,35){(i)}&\includegraphics[width=0.4\columnwidth]{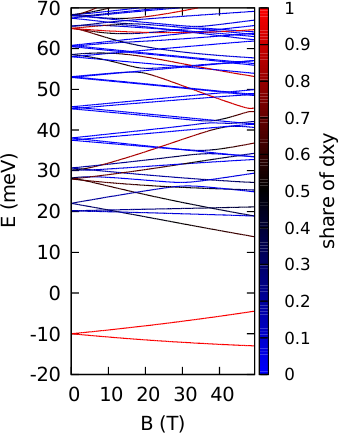}\put(-75,35){(j)}&\includegraphics[width=0.4\columnwidth]{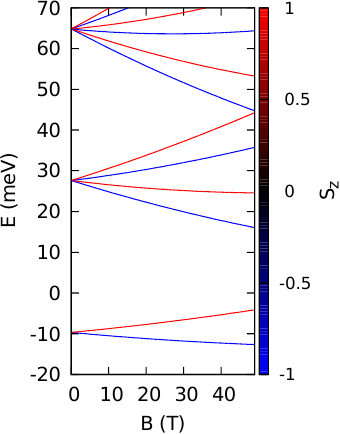}\put(-75,35){(k)}
&\includegraphics[width=0.4\columnwidth]{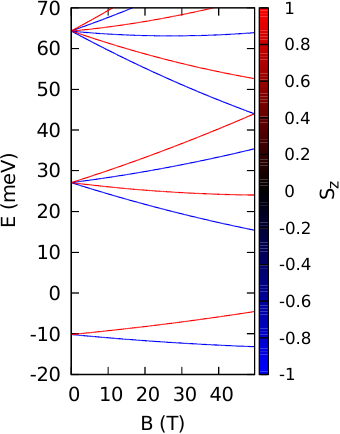}\put(-75,35){(l)}
\\
\end{tabular}
\caption{
Single-electron spectra as functions of the perpendicular magnetic field $B$, for parabolic confinement with energy $\hbar\omega_0 =9.344$meV (a-d),
$\hbar\omega_0 =18.689$ meV (e-h), and $\hbar\omega_0 =37.378$ meV (i-l). The left column of plots 
(a,e,i) presents the results of the exact Hamiltonian (\ref{eq:Hamiltonian_real}). The second column (b,f,j) displays the share of d$_{xy}$ orbitals
with the wave functions of the states within the Hamiltonian (\ref{eq:Hamiltonian_real}). The third column (c,g,k) shows the spectra as obtained with the d$_{xz}$ and d$_{yz}$ orbitals excluded from the basis, still with Hamiltonian (\ref{eq:Hamiltonian_real}). 
The last column (d,h,l) shows the results of the effective Hamiltonian (\ref{eq:Hamiltonian_real_xy}) reduced to the $d_{xy}$ orbitals.}
\label{singlespectra}
\end{figure*}

\subsection{Integration of the two-electron problem}
We consider the case of two electrons confined in QD embedded in 2DEG at the (001)-LAO/STO interface, described by the Hamiltonian 
\begin{equation}
\hat{H}_2=\hat{H}(1)+\hat{H}(2)+\frac{e^2}{4\pi\epsilon_0\epsilon r_{12}},
\end{equation}
where $\hat{H}$ is the single-electron Hamiltonian (\ref{eq:Hamiltonian_real}) or in the simplified model reduced to the $d_{xy}$ band, $\hat{H}=\hat{H}^{eff}_{xy}$, see Eq.~(\ref{eq:Hamiltonian_real_xy}). 
We take the dielectric constant $\epsilon=100 \epsilon _0$ which is the upper limit of the electric field dependence of $\epsilon$ given in Ref.~\onlinecite{Gariglio2015}. Although the dielectric constant of STO is known to be significantly dependent on the electric field and temperature, the assumed value is justified near the interface where the triangular quantum well is created and the electric field is large - \textcolor{black}{see the Appendix.}
\begin{table*}
\begin{tabular}{c|c|c|c|c|c|c|c|c|c}
$n$ & $E+\Delta E$ & $\Pi$ & $|eFx_{1n}|$ (meV)  &$d_{xy}\downarrow$&$d_{xy}\uparrow$&$d_{xz}\downarrow$&$d_{xz}\uparrow$&$d_{yz}\uparrow$&$d_{yz}\downarrow$  \\ \hline
1&	-29.30 &-1&	0&	0.99493&	0.00007&	0.00085&	0.00165&	0.00085&	0.00165 \\
2&	-27.57 &1&	0.00382 &	0.00010&	0.99444&	0.00186&	0.00087&	0.00186&	0.00087\\
3&	-12.72 &1&	1.866&	0.98711&	0.00019&	0.00412&	0.00223&	0.00412&	0.00223\\
4&	-11.00&-1&	0&	0.00041&	0.98529&	0.00250&	0.00465&	0.00250&	0.00465\\
5&	-8.81 &	1&1.859 &	0.97891&	0.00053&	0.00783&	0.00245&	0.00783&	0.00245\\
6&	-7.08&-1&	0&	0.00039&	0.97977&	0.00313&	0.00679&	0.00313&	0.00679\\
7&	3.67&-1&	0& 0.95177&	0.00103&	0.01921&	0.00438&	0.01921&	0.00438\\
\end{tabular}
\caption{The energy levels (second column) for $\hbar\omega_0=18.689$ meV at $B=(0,0,10)$T. The 3rd column gives the parity $\Pi$ eigenvalue.
The dipole matrix element with the ground state and the $n$ -th eigenstate is given in the fourth column for an electric field of 1mV/nm. The subsequent columns provide the
contributions of the atomic spin orbitals to the Hamiltonian (9) eigenstates. }
\end{table*}

The problem is solved in the basis of antisymmetrized products of the single-electron eigenfunctions, i.e. in the exact diagonalization approach applied for an electron pair.
For $H\Psi_q=E_q\Psi_q$, the single-electron eigenfunction is spanned over the $3d$ spin-orbitals of a Ti ion
\begin{eqnarray}
\Psi_q(x,y,\sigma)&=&\sum_j a_j^q d_j(x,y,\sigma)\nonumber \\ &=&\sum_{r_j,o_j,s_j} a_j^q d_{r_j,o_j}(x,y)S_{s_j}(\sigma),
\end{eqnarray}
where the summation runs over the position of ions $r_j$, orbitals $o_j$ on the ion and the z-component of the spin indexed by $s_j$, and 
$S$ is the spin-up or spin-down eigenstate.
In the sums $j$ is equivalent to the triple of indeces $(r_j,o_j,s_j)$ and $d_{r_j,o_j}$ is one of the 3d orbitals localized on the ion position $r_j$.

Evaluation of the matrix elements of the two-electron Hamiltonian requires determination of the Coulomb integrals
\begin{eqnarray}
&&I_{q_1q_2q_3q_4}=\langle \Psi_{q_1}(1)\Psi_{q_2}(2)|\frac{1}{r_{12}}|\Psi_{q_3}(1)\Psi_{q_4}(2)\rangle \\ \nonumber &=& 
\sum_{j_1,j_2,j_3,j_4}(a_{j_1}^{q_1}a_{j_2}^{q_2})^*a_{j_3}^{q_3}a_{j_4}^{q_4}\langle d_{j_1}(1) d_{j_2}(2)|\frac{1}{r_{12}}|d_{j_3}(1) d_{j_4}(2)\rangle.
\end{eqnarray}
The integral over the spin-orbitals that appears in the sum is calculated based on the formula
\begin{eqnarray}
&&\langle d_{j_1}(1) d_{j_2}(2)|\frac{1}{r_{12}}|d_{j_3}(1) d_{j_4}(2)\rangle\nonumber = \\&&\delta(r_{j_1},r_{j_3})\delta(r_{j_2},r_{j_4}) \delta(s_{j_1},s_{j_3})\delta(s_{j_2},s_{j_4}) \times \nonumber \\&&
\bigg[ \left(1-\delta(r_{j_1},r_{j_2})\right)\frac{1}{|r_{j_1}-r_{j_2}|}\delta(o_{j_1},o_{j_3})\delta(o_{j_2},o_{j_4})
+ \nonumber \\&& \delta(r_{j_1},r_{j_2}) \varepsilon(o_{j_1},o_{j_2},o_{j_3},o_{j_4}) \bigg ],
\label{CI:integral}
\end{eqnarray}
where 
\begin{equation}
\varepsilon(o_{j_1},o_{j_2},o_{j_3},o_{j_4})=\langle d_{j_1}(1)d_{j_2}(2)|\frac{1}{r_{12}}|d_{j_3}(1)d_{j_4}(2) \rangle
\end{equation}
is the integral for the four orbitals localized on the same ion.
The Kronecker deltas in the second line in Eq.~(\ref{CI:integral}) introduce the two-center approximation \cite{twocenter} and the orthogonality of the spin states. The third line is responsible for the contributions to the integral 
with the first and second electrons occupying different ions. In this term we assume that the Coulomb potential changes slowly allowing us to use the orthogonality of the orbitals. The last line of Eq. (25) is responsible for the Coulomb integrals over the same ion with electrons occupying various orbitals -- that is denoted as $\varepsilon$ integral. This integral is evaluated using the Monte-Carlo technique. For orbitals numbered as $d_{xy}=N \exp(-Z^*r/3)xy\rightarrow 1$, $d_{xz}=N \exp(-Z^*r/3)xz\rightarrow 2$, $d_{yz}=N \exp(-Z^*r/3)yz\rightarrow 3$, using the hydrogen-like 3d orbitals with the normalization factor $N$ and the effective atomic number $Z^*=3.65$ given by the Slater rules for the Ti orbitals, the non-zero values of the on-site Coulomb integral are: $\varepsilon(i,i,i,i)=0.336$, $\varepsilon(i,j,i,j)=0.306$,
and $\varepsilon(i,j,j,i)=0.015$ (for $i\neq j$) in the atomic units. The remaining 12 integrals with other sequence of the orbitals are zero due to negative parity of the integrated functions.

In the calculations, we use up to 50 lowest-energy single-electron states that produce 1225 Slater determinants as a basis for the two-electron problem.

\subsection{Time evolution}

For discussion of spin dynamics in the external electric field, we assume a periodic perturbation of the potential $V_{AC}(t)=-eFx\sin(2\pi \omega t)$. 
We solve the Schr\"odinger equation with the time-dependent Hamiltonian $\hat{H}_t=\hat{H}+V_{AC}(t)$. The solution is obtained on the basis of time-independent Hamiltonian eigenstates,  $\Psi(t)=\sum_m c_m(t) \exp\left(-iE_nt/\hbar\right)| m\rangle$   with $\hat{H}|m\rangle=E_m|m\rangle$.  Upon substitution of this wave function to the Schr\"odinger equation followed by a projection on the $\langle n|$ state we obtain a system of equations for $c_n(t)$, 
\begin{eqnarray}i\hbar c'_n(t)=-eF\sum_m&&c_m(t)\exp\left[i\left(E_n-E_m\right)t/\hbar\right] \nonumber \\ 
&&\times \sin(2\pi\omega t)\langle n|x|m\rangle,
\end{eqnarray}
 that is solved using the Crank-Nicolson scheme.
 In the initial condition we take $c_n(t=0)=\delta_{n,1}$.
 We monitor the maximal occupation ($\max |c_n(t)|^2)$
of the excited states during the simulation that covers a time interval of $5$~ns.

The present model assumes that the quantum dot is a closed system subject to external fields, hence no decay and dephasing mechanisms are taken into account.  In quantum dots the decay or relaxation of the system from the excited state to the ground-state with opposite spin occurs with electron-acoustic phonon coupling in presence of the spin-orbit interaction. In terms of the EDSR experiments the problem was studied in Ref.\cite{NSZPRB}. The relaxation may induce off-resonant transition to a third state with a lower energy than the couple participating in the EDSR. This situation corresponds in particular to the double quantum dot, when the states participating
in the EDSR are two spin states of charge configuration corresponding to separated
charges and the third state -- lower in the energy -- with both electrons in the deeper dot.
In the present work, with a single quantum dot considered there is no the third state
of a lower energy that could appear in the time evolution. In the simulations described in the manuscript the ground-state is always the initial state of the EDSR process.  

The main source of the spin decoherence in III-V or Si quantum dots is the
coupling of the electron spins with the nuclear spin-bath via the hyperfine interaction.
In terms of the EDSR experiments this decoherence leads to a reduction of the
spin-oscillations as a function of time \cite{Noiri}.
STO-based 2DEGs are expected to be characterized by a smaller 
of decoherence, as the Hyperfine Interaction (HFI) with the nuclear spin bath is intrinsically low in STO \cite{Ferrenti} since the electron bands at the Fermi level are built of 3d atomic orbitals  which vanish at the nuclei. 
Besides the spin-decoherence the evolution of the states can be perturbed by the charge nosite.
However, the latter is  mitigated by a large dielectric constant.

\section{Results}
\subsection{Single confined electron}

The Hamiltonian (\ref{eq:Hamiltonian_real}) in the basis limited to the d$_{xy}$ orbitals is equivalent to a single-band Hamiltonian with an isotropic electron effective mass of $m=\frac{\hbar^2}{2a^2t_l}=0.286m_0$, where $a=0.39$~nm is the lattice constant. 
Since the electrostatic confinement is typically parabolic near its minimum,
we have considered single-electron spectra for the external potential in the form $V(x,y)=\frac{1}{2} m \omega_0^2 r^2$ with
$\hbar \omega_0=9.344$ meV, $18.689$ meV, and $37.378$ meV, defining the range of a low, moderate and strong confinement. 
The spectra for a single confined electron are given in Fig. \ref{singlespectra}, for the case of the magnetic field applied perpendicular to 2DEG.

For the lowest confinement energy $\hbar\omega_0$ [Fig. \ref{singlespectra}(a)] the share of $d_{xy}$ orbitals [Fig. \ref{singlespectra}(b)]  in the low-energy part of the spectra is close to $1$ and the basis limited to d$_{xy}$ orbitals [Fig. \ref{singlespectra}(c)] provides nearly exact results in the low-energy range.
For $\hbar \omega_0=18.689$~meV [Fig. \ref{singlespectra}(e,f)] the states with low contribution of d$_{xy}$ orbitals appear already about $40$~meV above the ground state and the limited basis [Fig. \ref{singlespectra}(g)] produces results close to exact ones only for a few lowest-energy states. 
For the strongest confinement [Fig. \ref{singlespectra}(i-k)] only the two lowest energy states can be described with the limited model.
The last column of the plots [Fig. \ref{singlespectra}(d,h,l)] shows the results obtained with the Hamiltonian (\ref{eq:Hamiltonian_real_xy}) reduced to the $d_{xy}$ orbitals with SO coupling. Note that the simple limitation of the basis
to the $d_{xy}$ orbitals in Hamiltonian (\ref{eq:Hamiltonian_real}) [Fig. \ref{singlespectra}(c,g,k)] excludes all spin-orbit coupling interactions due
to the absence of direct coupling between the  $d_{xy,\uparrow}$  and $d_{xy,\downarrow}$ spin-orbitals. On the other hand the reduced model (\ref{eq:Hamiltonian_k_xy}) transfers the spin-orbit interactions originating from the coupling to $d_{xz}$ and $d_{yz}$ orbitals to the basis $d_{xy}$. Although the energy difference between the two approaches is small, the spin-orbit interaction in the model reduced to the $d_{xy}$ orbital is needed as it allows for control of the electron spin using the electric field. 
\begin{figure}
\begin{tabular}{cc}
\includegraphics[height=0.6\columnwidth]{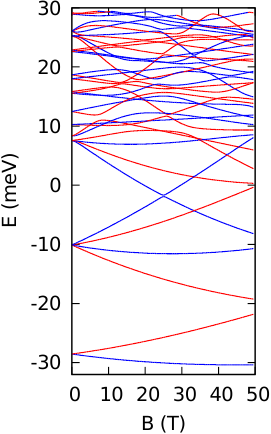}\put(-50,45){(a)}&\includegraphics[height=0.6\columnwidth]{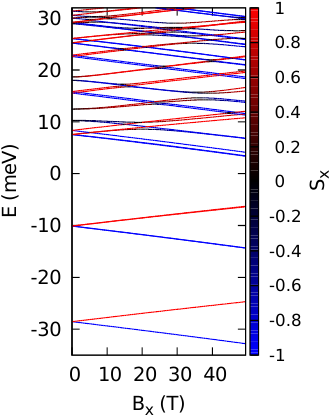}\put(-80,45){(b)}
\end{tabular}
\caption{(a) The single-electron spectrum of Fig. \ref{singlespectra}(e) ($\hbar\omega_0=18.688$ meV)
but plotted with the colours indicating the $\Pi$ parity of energy levels in a perpendicular magnetic field. The black color stands for the negative parity and the red
color for the positive parity.
(b) The single-electron spectrum with the same confinement energy but the external magnetic field
oriented in the $x$ direction. The colors indicate the sign of the $x$ component of the spin.}\label{parspin}
\end{figure}

Now, let us consider the transitions driven by the AC electric field oriented within the plane of confinement.
Direct transitions between the ground state and the excited states are governed by the values of the dipole matrix elements. The  values
  $x_{nm}=\langle n|x|m\rangle$  for  $n=1$ are listed in Table I for the magnetic field of $10$~T oriented perpendicular to the confinement plane.
The spin-flipping transition from the ground state to the first excited state $1\rightarrow 2$ has a small but non-zero matrix element. The states 1 and 2 have opposite parity, hence
the matrix element for $1\rightarrow 2$ transition is non-zero. The element is small
since the integration of the matrix element involves products of the spin-up
and spin-down contributions to both the states and their dominant spins are opposite.

The matrix elements for spin-conserving transitions to the 3rd and 5th states are about 500 times larger than $\langle 1 |x|2\rangle$.
A direct spin-flipping transition to the fourth state and the direct spin-conserving transition to the sixth state are forbidden.
There is a symmetry reason for the vanishing transition matrix elements to the 4th and 6th states. 
Note that the Hamiltonian (\ref{eq:Hamiltonian_real}) commutes with a generalized diagonal parity operator $\Pi=diag[P,-P,-P,P,-P,P]$,
where $P$ is the scalar parity operator $P\psi(r)=\psi(-r)$.
As a result, each of the components of the eigenfunctions has a definite -- even or odd -- scalar parity. It means that, for a given orbital, the $P$ parity of the spin-up and spin-down components is opposite. Moreover, for a given spin, the scalar parity of the $d_{yz}$, $d_{xz}$ components is the same and opposite to the parity of the $d_{xy}$ component.
The eigenvalues of the $\Pi$ parity for the lowest energy levels are listed in Table I and the spectrum with parity marked by colors
is plotted in Fig. \ref{parspin}(a).
The fourth and sixth states have the same parity as the ground state leading to the vanishing dipole matrix element for each of the six components in the scalar product $\langle 1|x|n\rangle$ with $n=4$ or $6$.

\begin{figure*}[htbp]
\begin{tabular}{lll}
\includegraphics[width=0.7\columnwidth]{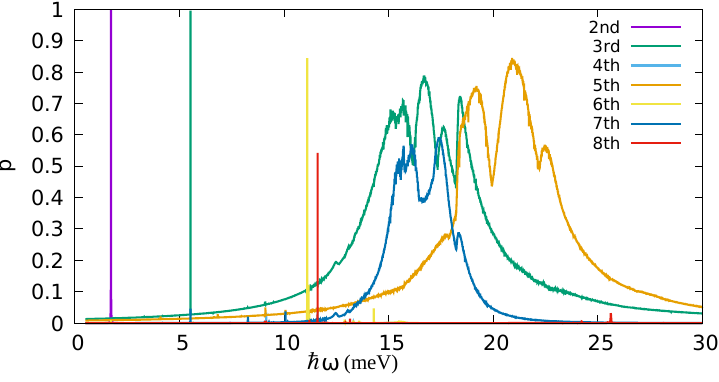}\put(-52,35){(a)}&
\includegraphics[width=0.7\columnwidth]{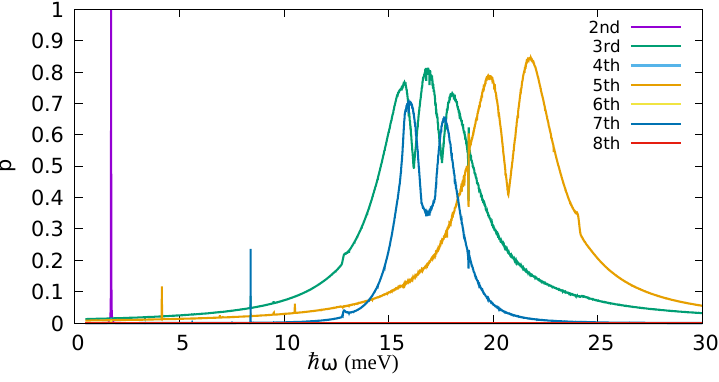}\put(-52,35){(b)}  \\
\includegraphics[width=0.7\columnwidth]{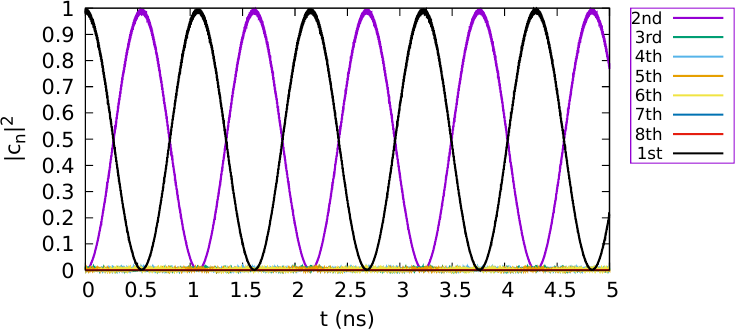}\put(-42,35){(c)} &  
\includegraphics[width=0.7\columnwidth]{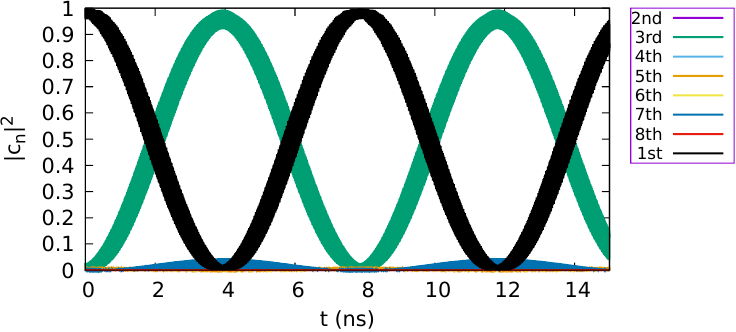}\put(-42,35){(d)} &
\includegraphics[width=0.7\columnwidth]{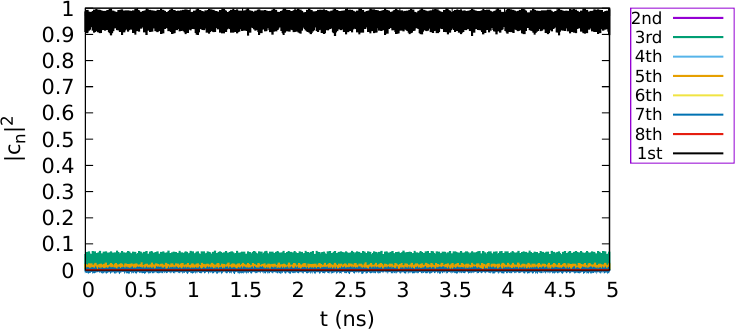}\put(-42,35){(e)} 
\end{tabular}
\caption{The results of the integration of the time-dependent Schr\"odinger equation for $\hbar\omega_0 =18.689$ meV, the perpendicular magnetic field of $10$~T with the AC electric potential, $V_{AC}(t)=-eFx\sin(\omega t)$. The simulation time is set to $5$~ns and the amplitude of the electric field to $1$~mV/nm. % (a-f) and 0.5mV/nm. 
In (a,b) we plot the maximal occupancy of the $n-th$ state defined as maximal $|c_{n}(t)|^2$ over the simulation time.
Panel (a) shows the results for the Hamiltonian (9) and panel (b) the results for the Hamiltonian (21), reduced to the $d_{xy}$ orbitals.
In (c-e) we plot the results obtained with the full Hamiltonian as in (a).
The lines show contributions of eigenstates as a function of time
for $\hbar\omega=1.727$ meV (c) -- where the peak for the direct $1\rightarrow 2$ transition is found in (a),  for $\hbar\omega=5.53$ meV  -- for the third-order transition peak $1\rightarrow 3$ of panel (a), and off-resonant excitation with $\hbar\omega=10$ meV.
In (c-e) we use the same colors for the same states as in (a,b). Additionally in (c,e) we
plotted the contribution of the ground state with the black line.
} \label{edsr1e}
\end{figure*}

Figure \ref{edsr1e} shows the maximum occupation for the driving AC field of 1 mV/nm with the full (a) and reduced (b) Hamiltonians, defined by Eqs.~(\ref{eq:Hamiltonian_real}) and (\ref{eq:Hamiltonian_real_xy}), respectively.
In Fig. \ref{edsr1e}(a) and (b), above the driving energy of 10 meV we can see wide overlapping maxima due to the allowed spin-conserving transitions to the 3rd and 5th state [cf. Table I]. In the same energy range, the maximum occupancy of the seventh energy level is also observed.
Note, however, that the direct transition $1\rightarrow 7$ is forbidden by the parity symmetry, but the energy difference $E_7-E_1$ is nearly equal to $2(E_3-E_1)$, so the transition to the 7th energy level occurs through transitions via the 3rd energy level $1\rightarrow 3\rightarrow 7$.
In Fig. \ref{edsr1e}(a) the narrow yellow and red peaks near 11 meV correspond to two-photon transitions to the 6th and 9th states, respectively, and appear at half the energy of the forbidden direct (i.e. single-photon) transition. The two-photon processes appear in the second-order time-dependent perturbation theory via a third state $m$ that intermediates the transition between the initial and final states.
In the second-order perturbation theory, the values of the coefficients in the wave function expansion are given by\cite{osika}
\begin{equation} c_n(t)=c_n(0)+c_n^1(t)+ c_n^2(t), \label{tdpte}\end{equation}
where
\begin{equation}
c_{n}^{(1)}(t)=\frac{-eF}{\hbar}x_{n1} \exp \left ( i\frac{(\omega_{n1}-\omega')t}{2} \right ) \frac{\sin(\frac{\omega_{n1}-\omega'}{2}t)}{(\omega_{n1}-\omega')}
\end{equation}
and
\begin{eqnarray}
c_{n}^{(2)}(t)&=&\frac{e^{2}F^{2}}{2i\hbar^{2}}\sum_{m}x_{nm}x_{m1}\nonumber \\
&&\bigg[ e^{i\frac{\omega_{nm}+\omega_{m1}-2\omega'}{2}t}\frac{\sin(\frac{\omega_{nm}+\omega_{m1}-2\omega'}{2}t)}{(\omega_{m1}-\omega')(\omega_{nm}+\omega_{m1}-2\omega')}\nonumber \\ && -e^{i\frac{\omega_{nm}-\omega'}{2}t}\frac{\sin(\frac{\omega_{nm}-\omega'}{2}t)}{(\omega_{m1}-\omega')(\omega_{nm}-\omega')}\bigg],
\label{eq:c_second}
\end{eqnarray}
with $\omega'=h\omega/\hbar$ and $\omega_{nm}=(E_n-E_m)/\hbar.$

The formula for $c_{n}^{(2)}(t)$ involves a sum of the matrix elements products $x_{nm}x_{m1}$ with a resonance at $2h\nu=E_{n}-E_{1}$ (hence the two-photon nomenclature), due to the expression in the second line of Eq.~(\ref{eq:c_second}). 
Although the direct (first-order) transition $1\rightarrow 6$ and $1\rightarrow 7$ is forbidden by the same parity of the initial and final states, the two-photon process
is still allowed due to nonzero values of matrix elements with $|m\rangle$ states of the opposite parity.

At a lower energy range of Fig. \ref{edsr1e}(a) we notice a peak at $h\omega=5.53$ meV for the transition to the third excited state. This in turn is a three-photon (third-order) process at the driving energy of $1/3$ of the energy difference between the energy levels (see Table I). The two-photon transition $1\rightarrow 3$ is forbidden, since the parity of any intermediate state $m$ will agree with the parity of states 1 or 3, thus making one of the dipole matrix elements in the product equal to zero - see Eq.~(\ref{eq:c_second}). 
 The contributions of the separate eigenstates as a function of time for
this transition is plotted in Fig. \ref{edsr1e}(d). The apparent large width of the
lines result from rapid oscillations of the contributions $|c_3|^2$ and $|c_1|^2$.
In Fig. \ref{edsr1e}(a) the peak does not exactly reaches 1. In Fig. \ref{edsr1e}(c) 
we can see that the contribution of the 7th energy level reaches maximum at 
the same moment as the contibution of the 3rd state, which limits the maximal value of the latter.

The lowest energy peak in Fig. \ref{edsr1e}(a) and (b) is the $1\rightarrow 2$ spin-flipping transition near the driving energy of 1.727 meV - compare with Table I. This is the EDSR spin-flip transition that we focus on in this work.
We find that at the resonance, for the amplitude of the electric field of $F=1$ mV/nm, the spin inversion time is 537.9 ps, and for the 
amplitude decreased by half the spin inversion time is twice longer -- just as for the Rabi oscillation
involving two states only. The time dependence of the contributions for this transition
is given in Fig. 4(d).
Note that indeed at the driving frequency for the spin-flip the higher-energy states have only a residual presence in the wave function, hence the transitions can be identified as the Rabi resonance.

The reduced model incorporating the spin-orbit effects to the effective $d_{xy}$ Hamiltonian (21), presented in Fig. \ref{edsr1e}(b), produces similar
results for the direct transitions, including the lowest-energy spin flip, and the structure of the wide maxima between 10 meV and 30 meV is similar.
The matrix elements for the reduced model are given in Table II. 
\begin{table}
\begin{tabular}{c|c|c|c|c|c|}\tiny
$n$& $E+\Delta E$&$\Pi$&$|eFx_{12}|$ (meV).  &$d_{xy}\downarrow$&$d_{xy}\uparrow$  \\ \hline
1&   17.404& -1 & 0 &    0.9998&      0.0002   \\
2&   19.139&  1&  0.0054&  0.00029&  0.99971      \\ 
3&   34.157&  1&  1.80&  0.99960&      0.0004   \\
4&   35.899&  -1 & 0 &  0.00049&  0.99951       \\
5&   38.198&  1&  1.880&  0.99951&      0.00049   \\
6&   39.922&  -1& 0&  0.00058&  0.99942       \\
7&   50.901&  -1& 0&  0.00069&  0.99931       \\
\end{tabular}
\caption{Same as Table I only for the $d_{xy}$-reduced Hamiltonian (21).}
\end{table}

\begin{figure}
\begin{tabular}{c}
\includegraphics[width=0.45\columnwidth]{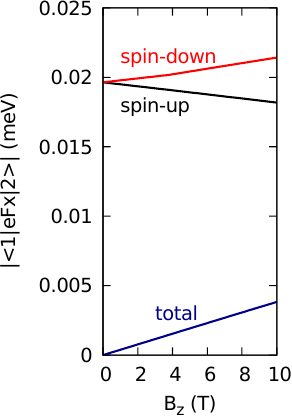}\put(-22,30){(a)}
\includegraphics[width=0.4\columnwidth]{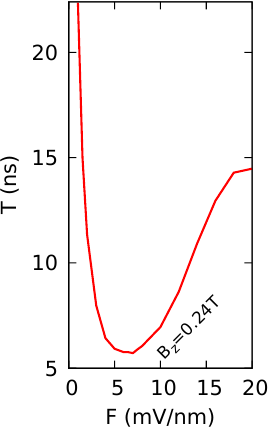}\put(-22,30){(b)}\\
\includegraphics[width=0.75\columnwidth]{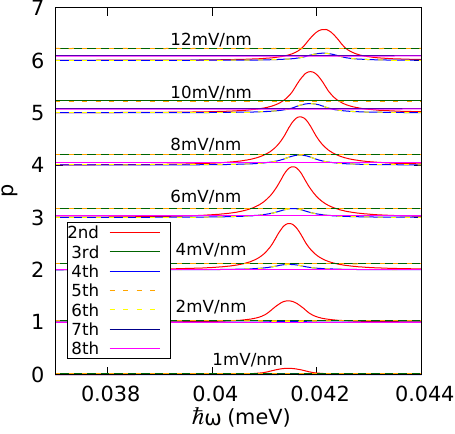}\put(-42,35){(c)}\\
\end{tabular}
\caption{The results for the single electron and confinement energy $\hbar\omega_0=18.688$ meV.
(a) The absolute value of the dipole matrix element for AC field of $F=1$~mV/nm (black line), and the absolute values
of the contributions to the matrix element from the spin-up and spin-down components. These contributions have opposite sign
and exactly cancel each other at $B=0$.
(b) The spin-flip time as a function of the AC amplitude for $B_z=0.24$ T, where the energy splitting between the 
ground (spin-down) and the first excited state (spin-up) is equal $\sim 0.0414$ meV that corresponds to the microwave frequency of $\sim 10$ GHz.
(c) The maximal occupancy of the lowest-energy levels for the time evolution starting from the ground-state and lasting 5ns. The AC field amplitude increases from
the bottom to the top. The subsequent plots for higher amplitudes are shifted by +1 each.
} \label{EDSR_single_electron}
\end{figure}

The estimated spin-flip time at the resonance for the amplitude of 1mV/nm is $379.81$~ps. Note, however, that the two-photon transitions to the sixth and eighth states as well as the three-photon transition to the third state of Fig. \ref{edsr1e}(a) are missing as these transitions occur via the state with the $d_{xz}$ and $d_{yz}$ component. Moreover, a narrow peak corresponding to the transition to the seventh excited state at 8.4 meV in Fig. \ref{edsr1e}(b) is pronounced but missing in Fig. \ref{edsr1e}(a). In summary, the reduced model does not exactly reproduce the results of the full model for the higher-order transition processes that involve intermediate states between the initial and final state in terms of the time-dependent perturbation theory. The spectrum of the reduced Hamiltonian misses part of the higher-energy states, which thus cannot assist in the transitions as the intermediates. Importantly, the EDSR transition between the lowest states with opposite spins is correctly captured by this simplified one band model.
\begin{figure*}
\begin{tabular}{cccc}
\includegraphics[width=0.4\columnwidth]{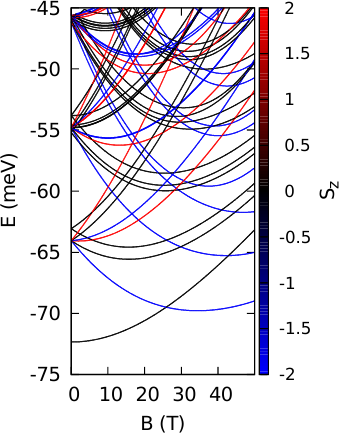}\put(-75,35){(a)}&\includegraphics[width=0.42\columnwidth]{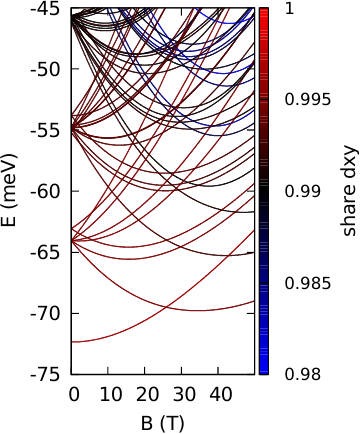}\put(-75,35){(b)}&\includegraphics[width=0.4\columnwidth]{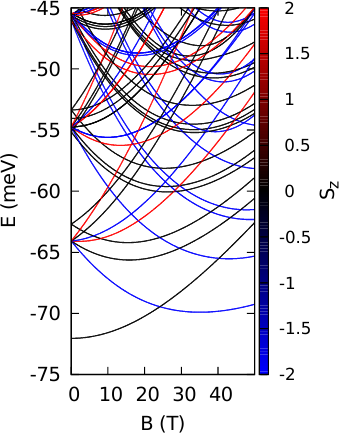}\put(-75,35){(c)}&\includegraphics[width=0.4\columnwidth]{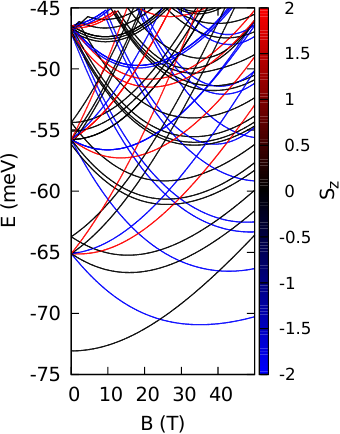}\put(-75,35){(d)}\\
\includegraphics[width=0.4\columnwidth]{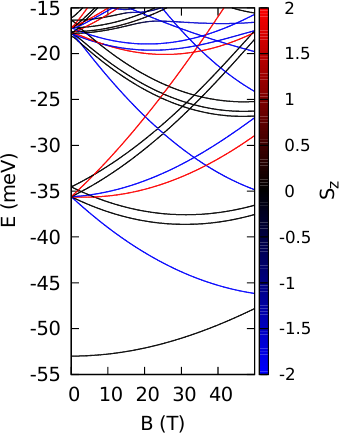}\put(-75,35){(e)}&\includegraphics[width=0.4\columnwidth]{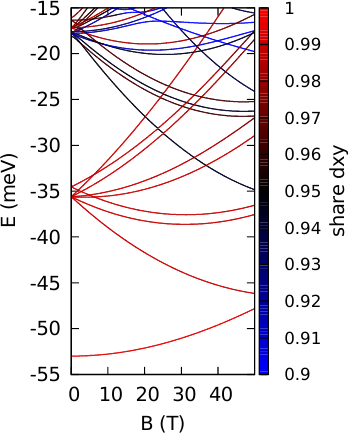}\put(-75,35){(f)}&
\includegraphics[width=0.4\columnwidth]{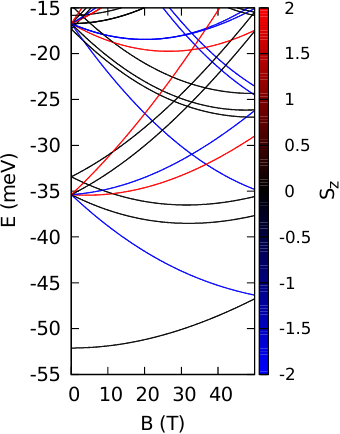}\put(-75,35){(g)}&\includegraphics[width=0.4\columnwidth]{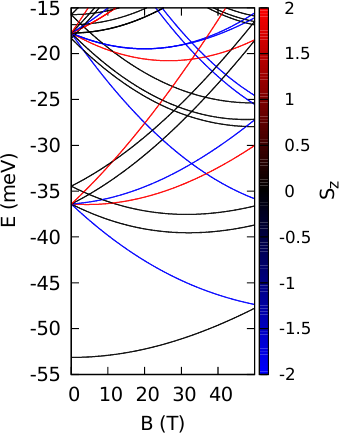}\put(-75,35){(h)}\\
\includegraphics[width=0.4\columnwidth]{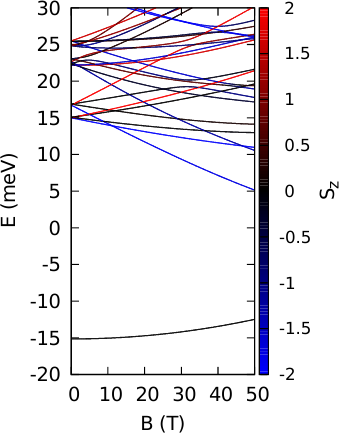}\put(-75,35){(i)}&\includegraphics[width=0.4\columnwidth]{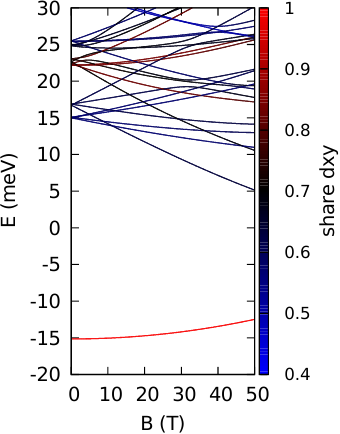}\put(-75,35){(j)}&\includegraphics[width=0.4\columnwidth]{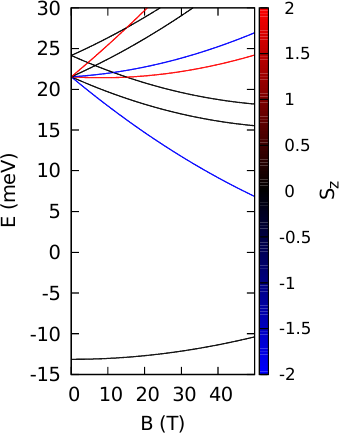}\put(-75,35){(k)}&\includegraphics[width=0.4\columnwidth]{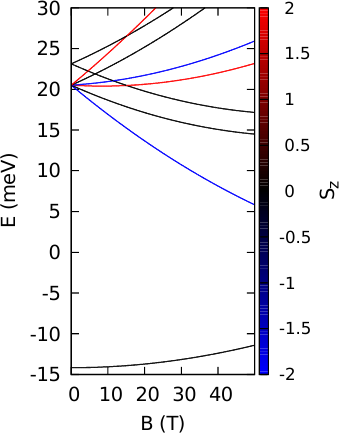}\put(-75,35){(l)}
\end{tabular}
\caption{Energy spectra for a confined electron pair $\hbar\omega_0 =9.344$meV (a-d),
$\hbar\omega_0 =18.689$ meV (e-g), and $\hbar\omega_0 =37.378$ meV (h-l). 
The first column of the plots shows the results of the full model Hamiltonian (\ref{eq:Hamiltonian_real}). The colour on the first, third and fourth (second) columns corresponds to the z component of the total spin (share of $d_{xy}$ orbitals). The third column shows the spectra for the Hamiltonian (\ref{eq:Hamiltonian_real}) but with the basis limited to d$_{xy}$ orbitals. The last column presents the results for $d_{xy}$-reduced Hamiltonian (\ref{eq:Hamiltonian_real_xy}), with the SO coupling effects integrated into the $d_{xy}$ band. 
}\label{twosp}
\end{figure*}

Finally, our calculations demonstrated that spin-flip from the ground state is only possible with the presence of both the atomic spin-orbit and the Rashba couplings. The hopping Rashba interaction is diagonal in spin and cannot drive spin transitions by itself. On the other hand, the atomic spin-orbit coupling
does not couple the $d_{xy}$ the spin-up and spin-down orbitals which dominate in the lowest-energy spectrum. Therefore, only a mutual presence of both spin-orbit interactions opens the way for spin flips.

The manipulation of the spin with an electric field is also possible for the magnetic field oriented
within the plane of confinement. Figure \ref{parspin}(b) shows the energy spectrum as a function of $B_x$. 
The second and third excited energy levels in Fig. \ref{parspin}(b) are nearly two times degenerate with a splitting of about 0.01 meV at $B_x=10$~T. At the scale of the Figure the lifting of the degeneracy
is not resolved. 
 For $B_x=10$ T and the amplitude of the electric field of 1mV/nm, the matrix element for the lowest-energy spin-flip transition induced by the AC electric field oriented in the $x$ direction is 3.7$\mu$eV. On the other hand, this matrix element is 0 for the AC field oriented in the $y$ direction. 
 The role of the orientation of the AC field and the external magnetic field is characteristic for the Rashba 2D interaction $H_R=\alpha_R (k_y\sigma_x-k_x\sigma_y)$ \cite{Rashba,Rashba2003}, which translates the motion in the $y$ ($x$) direction into an effective magnetic field oriented along the $x$ ($y$) axis~\cite{Bednarek}. The external magnetic field $(B_x,0,0)$ polarizes the spin parallel or antiparallel to the $x$ direction. Then the motion induced by the AC electric field parallel to the $x$ direction induces the $y$ component of the magnetic field that induces the spin-flips~\cite{Bednarek}. On the other hand, the AC field oriented in the $y$ direction produces an effective magnetic field oriented parallel or antiparallel to the external magnetic field. It can only affect the spin-splitting energy but does not couple the states with spins oriented in opposite directions along the $x$ axis.

\begin{table*}
\begin{tabular}{c|c|c|c|c|c|c|c|c|c}
$n$& $E+\Delta E$&$\Pi$&$|-eFx_{1n}|$ (meV).  &$d_{xy}\downarrow$&$d_{xy}\uparrow$&$d_{xz}\downarrow$&$d_{xz}\uparrow$&$d_{yz}\uparrow$&$d_{yz}\downarrow$  \\ \hline
  1& -52.276749&-1& 0&     0.995  &    0.994  &    0.00279 & 0.00252&  0.00279 & 0.00252\\
  2& -39.169408&-1& 0&      1.982&     0.000& 0.00484&  0.00381&  0.00484&  0.00381\\  
  3& -37.100982&+1& 0.0097&     0.991&      0.990&      0.00463&  0.00461&  0.00463&  0.00461\\
  4& -36.026479&+1& 2.63&      0.991&      0.990&      0.00462&  0.00457&  0.00463&  0.00457\\
  5& -35.032511&-1& 0&     0.001&   1.979&      0.00444&  0.00538&  0.00444&  0.00538\\
  6& -34.466481&-1& 0 &      1.972&      0.001&  0.00927&  0.00409&  0.00927&  0.00409\\
  7& -32.389505&+1& 0.0149&     0.986&      0.987&      0.00727&  0.00607&  0.00727&  0.00607\\
  8& -31.313806&+1& 2.62&     0.987 &     0.987&      0.00718&  0.00613&  0.00718&  0.00613\\
%  9& -30.312446&-1&      0.55213157E-03   1.9728096      0.52008721E-02  0.81182704E-02  0.52008721E-02  0.81182704E-02 
%  10&-23.064632&+1&       1.9518925      0.10411491E-02  0.17957738E-01  0.55754297E-02  0.17957738E-01  0.55754297E-02
%  11&-21.502877&      0.98273684      0.97641418      0.98006992E-02  0.10623790E-01  0.98006992E-02  0.10623790E-01  
%  12&-21.011984&      0.97635555      0.97136148      0.13132887E-01  0.13008598E-01  0.13132887E-01  0.13008598E-01  
%  13&-19.864294&      0.98257725      0.97938763      0.95173742E-02  0.95001853E-02   0.95173742E-02  0.95001853E-02
\end{tabular}
\caption{Results for two electrons, confinement energy $\hbar \omega _0=18.688$ meV and the vertical magnetic field of $12$~T. The results are organized as in Table I for a single electron.}
\end{table*}

The analysis given above describes the spin-flipping transitions for the energy difference between the low- and high-spin states of a few meV. Note, however, that EDSR experiments~\cite{Nadj-Perge2012} are usually performed with the AC field applied to the gate electrodes in a microwave range of about $10$~GHz which corresponds to the spin splitting energy of $\sim 0.041$~meV. 
For the single-electron spectrum with $\hbar\omega_0=18.689$ meV [cf. Fig.~\ref{singlespectra}(e-h)] this energy difference between the spin-down ground state and the spin-up first excited state corresponds to the magnetic field of the magnitude $B_z=0.24$~T for which the dipole matrix elements $\langle 1 |eFx| 2 \rangle$ are lower than at $B_z=10$~T. The dependence of $\langle 1 |eFx| 2 \rangle$ on the magnetic field is presented in Fig.~\ref{EDSR_single_electron}~(a). Note that the transition matrix element $\langle 1 | eFx|2\rangle$ is calculated by summation of the integrals over the six spin-orbital channels. 
We find that at $B=0$ the sum of components integrated over the spin-up orbitals is exactly opposite to the ones integrated over the spin-down orbitals [see Fig.~\ref{EDSR_single_electron}(a)]. In consequence, the transition matrix element at $B=0$ is exactly zero. 
As $B_z$ increases, the spin-down components are promoted by the spin Zeeman effect so that the contributions to the matrix element from both the spin channels become unequal which results in the non-zero value which changes a linear function of the external magnetic field [Fig.~\ref{EDSR_single_electron}(a)]. For the chosen magnetic field $B_z=0.24$~T and $F=1$ mV/nm we estimate the spin-flip time equal to $22.36$~ns  (compare with the spin-flip transition at $10$~T discussed in Section III.A which is $537.9$~ps only). The spin-flip time can be shortened by applying AC field of the larger amplitude. Note, however, that the increased amplitude of the AC field may lead to appearance of the higher energy states and changes the nature of the transition from the Rabi resonance to a more complex dynamics.
The spin flip time as a function of the amplitude is plotted in Fig.~\ref{EDSR_single_electron}(b) and the maximal occupancy
of the states in Fig.~\ref{EDSR_single_electron}(c). We find that the shortest spin-flip time of $5.71$~ns is found near $F\simeq 7$ mV/nm. Interestingly, the spin-flip time is an inverse function of the matrix element -- as in the Rabi oscillation -- only for $F$ below $4$~mV/nm. At larger $F$ the 3-rd and 5-th state (both spin-down) appear over the entire studied range of $\hbar\omega$ - see Fig.~\ref{EDSR_single_electron}(c). Importantly, the 4-th and 6-th states (both spin-up) appear within the range of the resonant spin-flip maximum, that indicates that the electron is first transferred to the 2-nd energy level and next strongly couple to higher spin-up energy states \cite{Sherman3}.  Note, that the spin-flip resonance in Fig.~\ref{EDSR_single_electron}(c) is black-shifted for a higher AC field amplitude. This effect is known as the Bloch-Siegert shift \cite{szirlej}.

\subsection{Two confined electrons}
\subsubsection{Spectra}
The energy spectra for the confined electron pair are diplayed in Fig. \ref{twosp},
for the exact Hamiltonian (\ref{eq:Hamiltonian_real}) (first three columns of plots) and the effective Hamiltonian (\ref{eq:Hamiltonian_real_xy}).
In the third column the basis was limited to solely $d_{xy}$ orbitals in Hamiltonian (\ref{eq:Hamiltonian_real}).
The efective Hamiltonian (fourth column) and the limited basis applied to the exact Hamiltonian (third column) produce similar results although with a small shift on the energy scale. The splitting between the energy levels that move parallel in the magnetic field is smaller in the exact 
Hamiltonian (the first column in Fig. \ref{twosp}) than in the approximate approaches (the last two columns in Fig. \ref{twosp}). The share of $d_{xy}$ orbitals in the low-energy states depends on the energy eigenvalue and the strength of confinement similarly as in the single-electron case (cf. Fig. 2).

\begin{figure} 
\begin{tabular}{c}
\includegraphics[width=0.8\columnwidth]{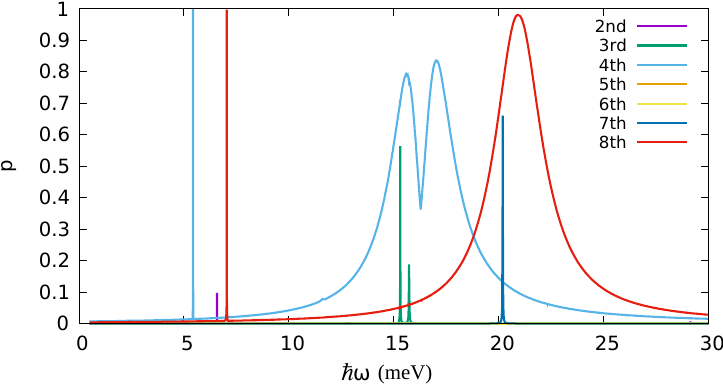}\put(-42,35){(a)}\\
\includegraphics[width=0.8\columnwidth]{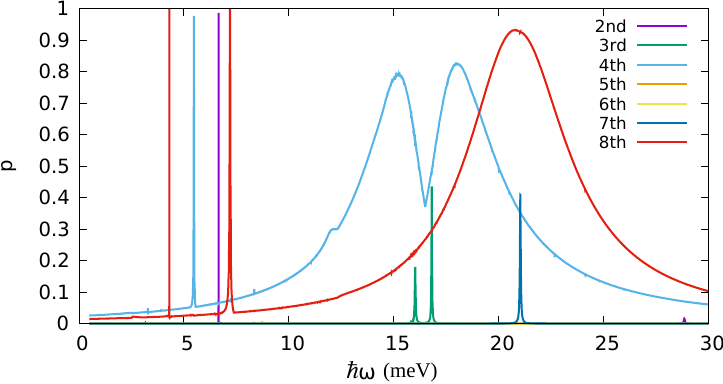}\put(-42,35){(b)}\\
\includegraphics[width=0.8\columnwidth]{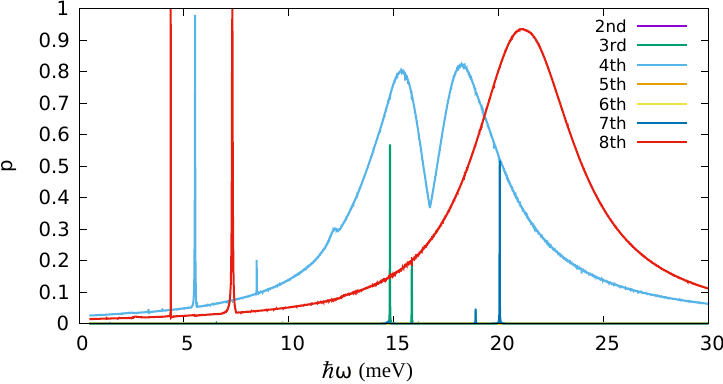}\put(-42,35){(c)}
\end{tabular}
\caption{
The results for two electrons, confinement energy $\hbar \omega _0=18.688$ meV and the vertical magnetic field of $12$~T.
Maximal occupation probability of excited states for the ground state as the initial one in the time evolution
lasting $5$~ns for AC potential $V_{AC}=-eFx\sin(h\omega t)$ with the amplitude (a) $0.5$~mV/nm and (b,c) $1$~mV/nm.
Panels (a) and (b) show the results of the full model while results for the effective Hamiltonian reduced to the $d_{xy}$  orbitals (\ref{eq:Hamiltonian_real_xy})
are given in (c).  The spin-flip transition $1\rightarrow 2$ in the two-photon process takes 4.7 ns for $F=1$~mV/nm.}
 \label{x11}
\end{figure}

In the following, we are mainly interested in the spin-flip between the spin singlet ground state and the first-excited triplet state with the spins polarized antiparallel to the external magnetic field.
The contributions of the orbitals, the parity, and the dipole matrix elements are given in Table III 
for $\hbar\omega_0=18.688$ meV and the perpendicular magnetic field of $12$~T.
In the table we see that the transition of the main interest $1\rightarrow 2$ is forbidden by the parity symmetry while the transitions are allowed to the pairs of even $\Pi$ parity states -- the 3rd and the 4th as well as the 7th and the 8th. Each pair is split by  $\sim 1$ meV and the energy levels of the pair move parallel when the magnetic field is changed [Fig. \ref{twosp}(e)]. In each pair, the transition to a lower-energy state has a much smaller probability than to a higher-energy one which can be explained as follow. In the basis limited to $d_{xy}$ orbitals (third column in Fig.~\ref{twosp}) the spatial and spin degrees of freedom separate due to the absence of the spin-orbit coupling effects. Then, the lower-energy state corresponds to the spin-triplet
with spin $z$ component equal to zero and the antisymmetric spatial wave function.
The higher-energy state corresponds to the symmetric spatial wave function
characteristic to the spin singlet.

The matrix elements for the transition from the singlet ground-state calculated with the $x_1+x_2$ operator are then zero due to the antisymmetry of the spatial part of the wave function with respect to the electrons interchange. The lifting of the separation between the spin and the space
by the spin-orbit interaction included in the complete basis opens the direct channel for the transition to the lower-energy level of each pair, but the matrix element for the lower-energy state is small.

 The results of the time evolution in $V_{AC}(t)$ field are plotted in Fig. \ref{x11} where, for the AC amplitude of $0.5$~mV/nm, we can see wide maxima corresponding to the direct transitions to the 4th and 8th states.
Besides the wide maxima we can observe a peak for the direct transition to the 7th state near $\hbar\omega=20$ meV.
Two narrow peaks corresponding to the transition to the 3rd state appear near $\hbar\omega=15$ meV and next we see a third order transition to the 8th state. The second order transition to this state is forbiden by the symmetry (cf. Table III).
To left of this peak there is the lower one due to the two-photon transition of our main interest: to the second energy level -- the one 
which corresponds to the flip of one of the spins from the ground-state for which
the direct transition is forbidden by the parity symmetry.

\begin{table}
\begin{tabular}{c|c|c|c|}
$n$&  $|eFx_{1n}|$,$V_0=2$meV& $|eFx_{1n}|$,$V_0=20$meV& $\leftarrow$ reduced mod.\\ \hline
1 & 0.1300 & 1.1726 &1.1298\\
2 & 0.0006 & 0.0039& 0.0013\\
3 & 0.0094 & 0.0058&0.0029\\
4 & 2.5800 & 2.0464&2.1145\\
5 & 0.0019 & 0.0094&0.0558\\
6 & 0.0011 & 0.0098&0.0042\\
7 & 0.0056 & 0.0054&0.0049\\
8 & 2.6800 & 3.0296&3.0039\\
\end{tabular}
\caption{The dipole matrix elements in meV for the two-electron system with the Gaussian perturbation
to the harmonic oscillator potential with confinement energy $\hbar \omega_0=18.688$~meV and the vertical magnetic field of $12$~T, for the amplitude of the AC electric potential $eF$ of $1$~mV/nm. The second (third) column gives the results for $V_0=2$~meV ($V_0=20$~meV). The 4th column lists the results for $V_0=20$ meV obtained with the reduced Hamiltonian.} 
\end{table}

\begin{figure}
\begin{tabular}{cc}
\includegraphics[width=0.5\columnwidth]{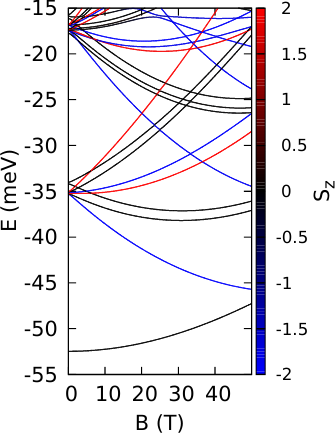}\put(-50,30){(a)}&
\includegraphics[width=0.5\columnwidth]{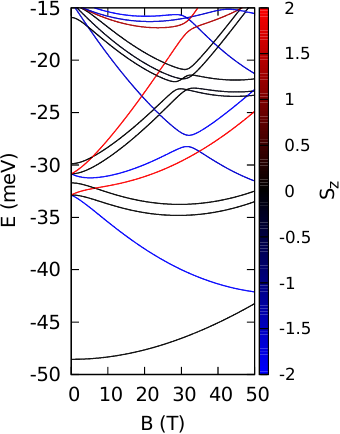}\put(-50,30){(b)}\\
\end{tabular}
\caption{The energy spectra for two-electrons and the confinement energy $\hbar \omega_0=18.688$ meV in presence of the off-center Gaussian perturbation with (a) $V_0=2$ meV and (b) $V_0=20$ meV.
} \label{zaburzenie}
\end{figure}

For the  amplitude of the AC field increased to $1$~mV/nm  [Fig. \ref{x11}(b)]  the spin-flip $1\rightarrow 2$ transition probability within the $5$~ns of the simulation
is increased to 98.8\%  via the second order two-photon process that takes $4.7$~ns. For comparison, for the AC field amplitude set at $0.5$~mV/nm the spin flip via the second order process takes $24.7$~ns 
and the transition to the triplet state is achieved with the probability of 99.645\%.
The increased fidelity of the spin-flip by reduction of the AC field amplitude is due to the reduction of the contribution of higher energy states
in the time evolution (cf. Fig. \ref{x11}(a) and (b)). Therefore, there is a trade-off between the transition time and the fidelity.
Note, that in the results for the effective Hamiltonian reduced to the $d_{xy}$ orbitals the peak due to the two-photon second-order spin-flipping 
transition to the second energy level is missing [Fig. \ref{x11}(c)]. Although the reduced model worked well
for the direct spin-inversion in the single-electron case, the second-order transition is missed out.
In fact a closer inspection of the results reveals the second-order peak but with a tiny magnitude of the order of $10^{-4}$,
which is outside the resolution of Fig. \ref{x11}(c). Already in the single-electron case we noticed
that the reduced model is not completely reliable for the description of the higher-order transitions, which
go via intermediate states, part of which are missing in the reduced Hamiltonian eigenspectrum.
\begin{figure}
\begin{tabular}{c}
\includegraphics[width=0.85\columnwidth]{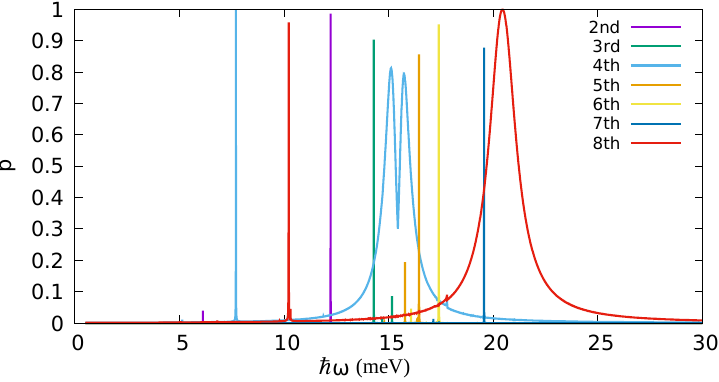}\put(-42,35){(a)}\\
\includegraphics[width=0.85\columnwidth]{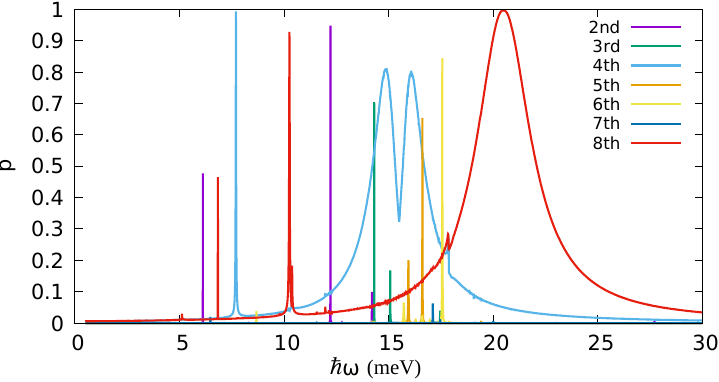}\put(-42,35){(b)}\\
\includegraphics[width=0.85\columnwidth]{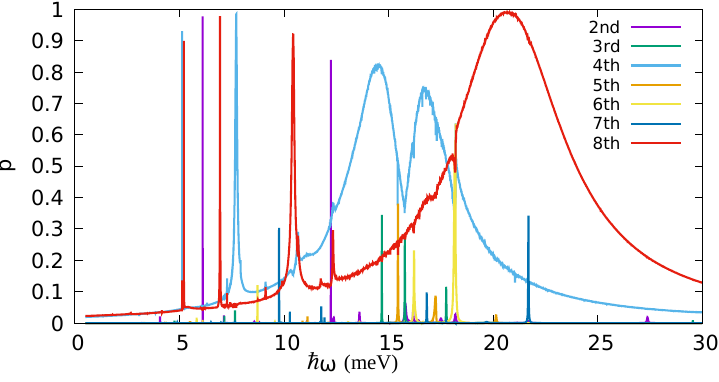}\put(-42,35){(c)}\\
\end{tabular}
\caption{The maximal occupation of the states for the simulation starting from the ground-state and lasting $5$~ns. Results for two electrons with Gaussian perturbation with $V_0=20$ meV.
} \label{zaburzenieedst}
\end{figure}

The two-photon process for the spin-flip transition $1\rightarrow 2$ is not very fast and the direct one is missing due to the dipole matrix element that vanishes due to the parity symmetry reason. 
One can try to speed-up the process by e.g. perturbation of the confinement potential
lifting its inversion symmetry and thus the parity selection rule.
In order to lower the symmetry we have placed a Gaussian perturbation $V_g=V_0 \exp\left(-\left((x-x_r)^2+y^2\right)\right/s^2)$, with $x_r=3$ nm and $s=2$ nm, to the parabolic confinement potential.
The matrix elements for $V_0=2$ meV and $V_0=20$ meV are listed in Table IV.
We see that the largest transition elements to 4th and 8th states are weakly changed by the Gaussian perturbation to the potential while 
the transition from the ground state to the first excited state, involving
a transition from zero to spin-polarized spin becomes nonzero in presence of the
perturbation.

\begin{figure}
\begin{tabular}{cc}
\includegraphics[width=0.5\columnwidth]{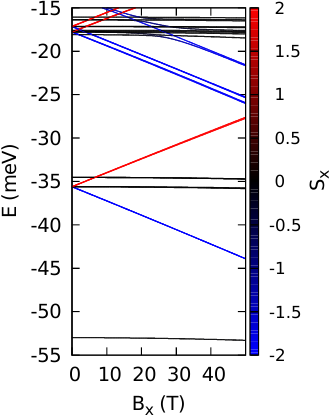}\put(-50,35){(a)}&
\includegraphics[width=0.52\columnwidth]{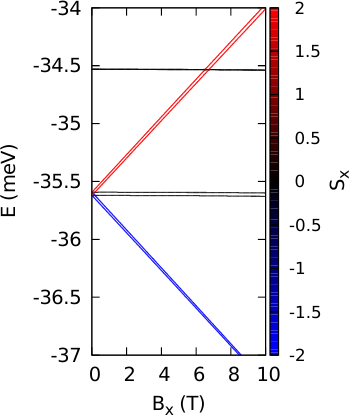}\put(-50,35){(b)}\\
\end{tabular}
\caption{The energy spectrum for two-electrons with $\hbar\omega_0=18.688$ meV 
in the magnetic field oriented parallel to the $x$ axis. Panel (b) is a zoom of (a). 
} \label{2ewbx}
\end{figure}

The two-electron energy spectra for the potential with the Gaussian perturbation are displayed in Fig. \ref{zaburzenie} for $V_0=2$~meV and $V_0=20$~meV. The Gaussian perturbation opens avoided crossings between states of the
same spin and the parity which is opposite for the symmetric potential. 
The transition spectra are plotted in Fig. \ref{zaburzenieedst} for $V_0=20$ meV and the AC field
amplitude of $0.25$~mV/nm (a), $0.5$~mV/nm (b) and $1$~mV/nm (c).
In each panel both the first-order and the second-order spin flipping transitions 
$1\rightarrow 2$ are observed. As the amplitude of the AC field increases, the second-order peak near $\hbar \omega=6$~meV grows up
but the first-order peak near $\hbar \omega=12$~meV decreases. The decrease of the single-photon transition is due to the widening
of the maxima related to transitions to the 6th and 8th states that compete in the evolution process
with the transition to the 2nd state. 
The spin-flip times for the second-order process for the AC amplitudes of 0.25, 0.375, 0.5, 0.75 and 1 mV/nm
are 39.2, 17.7, 10.2, 4.9 and 3.09 ns, respectively. The maximal occupation of the second energy level are 99.9\%, 99.6\%, 99\%, 98.7\% and 97.7\%, respectively.
For the first-order transition the spin flip times are: 2.25, 1.6, 1.4, 1.24, and 3.1 ns,
with the fidelity of the transfer to the 2nd excited state of 98.6\%, 96.9\%, 91.2\%, 90.6\% and 83.7\%. Remarkably, in this case
the spin-flip transition slows down when the amplitude is increased from 0.75 mV/nm to 1 mV/nm
that is due to participation of the other excited states in the time evolution besides the initial and the targeted one.
For both the second- and first-order transitions
the fidelity of the spin flip decreases with the AC amplitude due to leakage of the wave function to the higher-energy states. 
However, the fidelity of the transfer via the second-order processes is larger due to the lower background of
the other excited states in the lower energy range. 
\begin{table}
\begin{tabular}{c|c|c|c|}
$n$& $E+\Delta E$&$\langle S_x\rangle$&$|eFx_{1n}|$ ($\mu$eV)\\ \hline
1&    -53.009& -0.001 & 0  \\
2&   -37.574&  -1.993&  3.42 \\
3&   -37.563&  -1.990&  0.021 \\
4&   -35.632   &  -0.000 & 0.049 \\
5&   -35.602   &  -0.001   &12.228 \\
6&   -34.542 &   -0.000 & 4.630\\
7&   -34.541    & -0.001& 3737.6\\
8&   -33.686 &      1.992  &9.046\\
9&   -33.642    &   1.988 &0.028 \\
\end{tabular}
\caption{The energy levels, the average value of $S_x$, and the transition matrix element for $B_x=12$~T (see Fig. \ref{2ewbx})
with the AC field amplitude of $1$~mV/nm.} 
\label{t2ewbx}
\end{table}

\begin{figure}
\begin{tabular}{c}
\includegraphics[width=0.75\columnwidth]{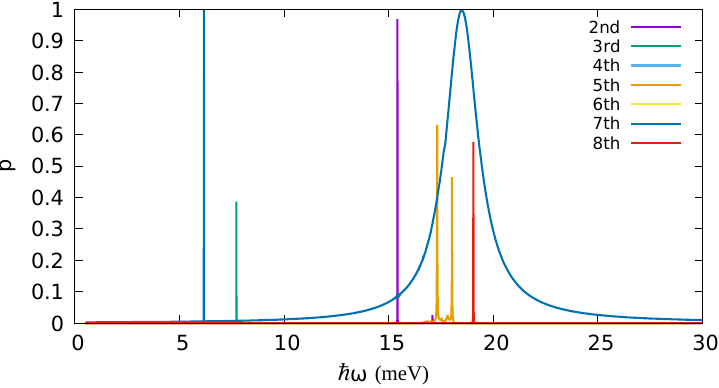}\put(-42,35){(a)}\\
\includegraphics[width=0.75\columnwidth]{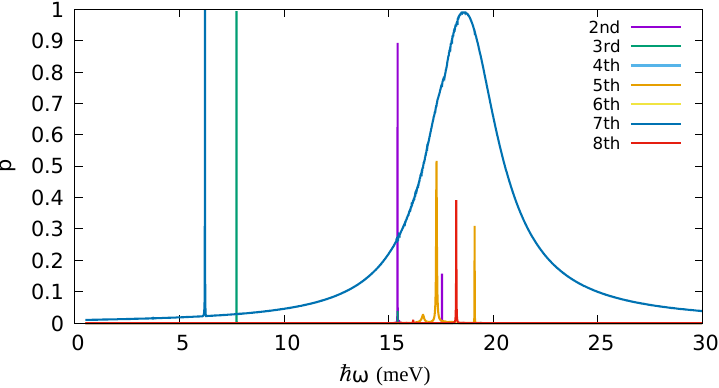}\put(-42,35){(b)}\\
\includegraphics[width=0.75\columnwidth]{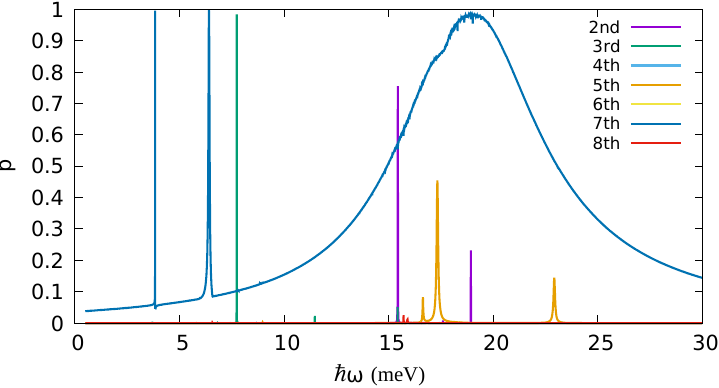}\put(-42,35){(c)}\\
\end{tabular}
\caption{The results for two electrons for confinement energy $\hbar\omega_0=18.688$ meV and the external magnetic oriented parallel 
to the $x$ axis, $B_x=12$ T. The maximal occupation of the states for the simulation starting from the ground state
and lasting $5$~ns is shown. The amplitude of the AC electric field oriented in the $x$ direction is (a) $0.25$, (b) $0.5$ and (c) $1$~mV/nm.
} \label{2ewxed}
\end{figure}

The first order spin-fliping transitions can also be observed for an ideally parabolic confinement potential but with the in-plane magnetic field that lifts the $\Pi$ symmetry.
In Fig. \ref{2ewbx}(a) we plotted the two-electron spectrum as a function of the magnetic field oriented along the $x$ axis
with the structure of the lowest-energy excited state enlarged in Fig. \ref{2ewbx}(b).
The value of the energies obtained at $B_x=12$~T with the average spin $x$ component and the 
transition matrix elements from the ground state are given in Table V.
At $B_x=12$~T all the excited states considered in Table V are nearly two-fold degenerate.
The magnetic field oriented in-plane does not produce the orbital effects present for the perpendicular
magnetic field hence the splitting is primarily due to the spin-orbit interaction.
The splitting of the excited energy levels is resolved in Fig. \ref{2ewbx}(b) with the exception of the 
two energy levels below -34.5 meV that in the absence of the SO interaction correspond to 
spin-singlet. All the other energy levels in Fig. \ref{2ewbx}(b) 
in the absence of SO coupling correspond to a spin-triplet. 
For each couple of the excited states the transition matrix elements from the ground-state is large
for one of the states of the couple and much smaller to the other - see Table V. In the table we see that the largest is the transition matrix element to an excited ''singlet'' state. 
The simulated spectra of excitations are plotted in Fig. \ref{2ewxed} for the AC electric field oriented along the $x$ axis with the amplitude (a) $0.25$, (b) $0.5$ and (c) $1$~meV.
The transition to the spin singlet -- the 7th state -- corresponds to the wide maximum, with the narrower
peaks due to the 3rd order process and 5th order process (c) for the largest AC amplitude. 
 We find that the spin-flip channels go through the first-order process
for the 2nd energy level and the second-order process for the 3rd energy level. 
 As the AC amplitude is increased the peak corresponding to
the second order process increases and the one due to the direct transition is reduced due to the 
strengthened presence of the 7th energy level in this excitation energy range. 
The spin-flip times for the AC amplitudes of 0.25, 0.5 and 1 mV/nm 
in the first order transition to the 2nd energy level are 2.4 ns, 1.1 ns, and 502 ps,
with the fidelity of the transfer of 96.9\%, 89.3\% and 75.6\%, respectively.
The corresponding numbers for the second-order transitions to the 3rd state and the listed amplitudes are: 11.67 ns, 3.03 ns and 845 ps 
and 99.64\%, 99.54\% and 98.48\%, respectively. 
\begin{figure}{}
\begin{tabular}{c}
\includegraphics[width=0.75\columnwidth]{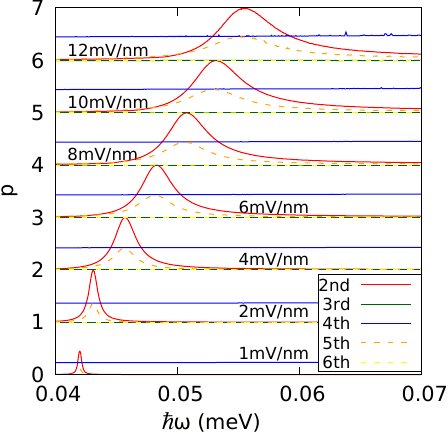}\put(-22,165){(a)}\\
\includegraphics[width=0.75\columnwidth]{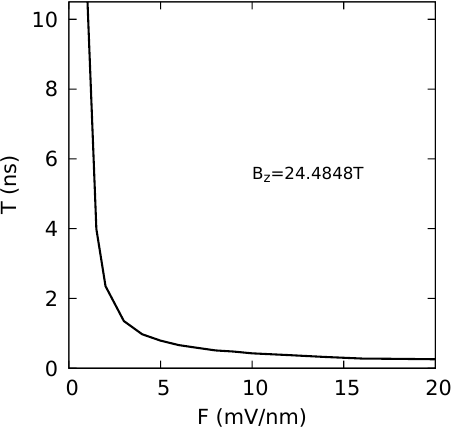}\put(-22,165){(b)}
\end{tabular}
\caption{The results for the two electrons and confinement energy $\hbar\omega_0=9.344$ meV at $B_z=25.4848$ T.
(a) The maximal occupancy of the lowest-energy levels for the time evolution lasting $5$~ns starting from the spin-singlet ground-state and AC field amplitude increasing from the bottom to the top. The subsequent plots for higher amplitudes are shifted by +1 each. (b) The singlet-triplet transition time as a function of the AC amplitude.
} \label{EDSR:two_electrons}
\end{figure} 

Similarly as in the single electron QD now, let us analyze what happens when $V_{AC}(t)$  is set in the GHz regime.
For a symmetric confinement potential and vertical magnetic field the singlet-triplet first-order transition is forbidden by the parity selection rule. To set the spin-flip AC field frequency in the second-order process to about $10$~GHz we need the singlet-triplet energy difference to be equal to about $2\times 0.0413$ meV. In the magnetic field range given in Fig.~\ref{twosp} this situation is observed only for the weakest confinement of $\hbar\omega_0=9.344$ meV, i.e. at $B_z=25.4848$ T, just below the singlet-triplet crossing in Fig.~\ref{twosp}(a). 
Fig.~\ref{EDSR:two_electrons}(a) shows the maximal occupancy of the lowest-energy levels for the $5$~ns time evolution in AC field as a function of the driving frequency. The black-shift for this second-order process
with the field amplitude is stronger than in the first-order process discussed above for the single-electron. The fourth excited state (spin-singlet) participates in the time evolution for all the considered amplitudes. The 5th excited state (spin-down polarized triplet) appears in the evolution near the resonance for the singlet-triplet transition.
The spin-flip time can be taken from $10.6$~ns (for $F=1$ mV/nm) to $0.22$~ns (for $F=20$~mV/nm). 
The transitions are faster than in the single-electron case discussed above but the magnetic field is by two-orders of the magnitude larger. 

{\color{black}
Finally, as the considered QD is created electrostatically by the top and bottom gate, another issue which can be regarded is the possible application of the AC electric field perpendicular to the 2DEG using those gates. Although this specific scenario was not addressed in the manuscript, in order to delve into this situation further, three issues need to be raised: (i) as stated in our manuscript, in the considered 2D model of QD, the transition rate between the states with opposite spins depends on the relative orientation between the magnetic and electric fields -  they need to be oriented in such a way that the effective SO magnetic field has a component perpendicular to the spin direction. For this reason, in our calculations, no transitions are observed when the magnetic field is directed along the y-axis. In the case of the perpendicular electric field, the effective SO field is perpendicular to the spin orientation for the out-of-plane magnetic field. It means that the transitions between the spin states oriented in-plane should have a similar character as described in the paper. (ii) The orbitals of d-electrons are spatially oriented due to the vertical electric field. For this reason, we would expect that $d_{xz/yz}$. orbitals could play a more significant role for the perpendicular field orientation, leading to the increase of the transition elements between those orbitals. (iii) The model considered in the paper is the projection of the real 3D Hamiltonian into 2D x-y space, assuming that electrons occupy the ground state related to the confinement in the z-direction. The transitions induced by a vertical AC field would involve excited states of the vertical quantization. Due to the nature of the 2DEG, these excitations should require a large amount of energy. The description of these processes is however beyond the scope of the present paper.}

\section{Summary and Conclusions}
We have studied a single and two electrons confined in a lateral quantum dot defined within the two-dimensional electron gas
on the (001)-oriented LAO/STO surface. For this purpose, we have developed a real space tight-binding Hamiltonian spanned by 3D orbitals of Ti. The reduced Hamiltonian integrating the spin-orbit coupling effects due to the $d_{xz}$ and $d_{yz}$ orbitals into an effective in-plane $d_{xy}$ band has been also derived and analyzed with respect to the full 3 bands model. 
We have analyzed the energy spectrum in a parabolic confinement and demonstrated that for a weakly confined systems the low-energy eigenstates
can be identified with the $d_{xy}$ orbitals. In this case the spectrum is close to the one of the harmonic oscillator with the electron effective mass
of $m=0.286 m_0$. For stronger confinement the states related to the orthogonal bands appear lower in the energy spectrum.

In the paper, we  have discussed the manipulation of the confined spin by external AC voltages in the context of the electric dipole spin resonance and
demonstrated that the spin-flip in the ground-state can be accomplished by a Rabi resonance with the transition time of the order of $0.5$~ns for
the amplitude of the AC field of the order of 1mV/nm. 
For the electron pair in the harmonic oscillator potential and the perpendicular magnetic field the singlet-triplet transition is forbidden by the 
parity symmetry. However, the spin-flip can still be obtained via a second-order, two-photon process that has a two-state Rabi character for low AC field amplitude. The parity selection rule excluding the single-order transition can be lifted by a perturbation of the external potential or in-plane orientation of the external magnetic field. In this case the first-order transition deviates from the Rabi oscillation due to the participation of higher energy singlet states in the time evolution. We have also found that fidelity of the transfer increases when lowering the  amplitude of the AC field and can reach  almost 97\% values.
Our results can be verified in EDSR experiments on LAO/STO QD~\cite{Jespersen2020} and demonstrate the possibility of quantum operation on oxides QDs
with high fidelity and fast control.

\section{ACKNOWLEDGEMENT}
This work was financed by the Horizon Europe EIC Pathfinder under the grant ``Spin-orbitronic quantum bits in reconfigurable 2D-oxides'' (IQARO), number 101115190.
Computing infrastructure PLGrid (HPC Centers : ACK Cfronet AGH) within computational grant no. PLG/2023/016317 was used.

\appendix*
\section{Dielectric constant at the LAO/STO interface.}
{\color{black}
It is widely known that the dielectric constant of strontium titanate strongly depends on the temperature and electric field, reaching values as high as $10000$. However, it is important to note that the electric field near the interface where 2DEG is created is very strong, leading to a reduction in the dielectric constant~\cite{Scopigno}. To support our assumption regarding the relatively low value of $\epsilon$ at approximately 100$\epsilon_0$, we conducted additional calculations to determine the dielectric permittivity profile at the LAO/STO interface 
\begin{figure}[!b]
\begin{tabular}{c}
\includegraphics[width=1.00\columnwidth]{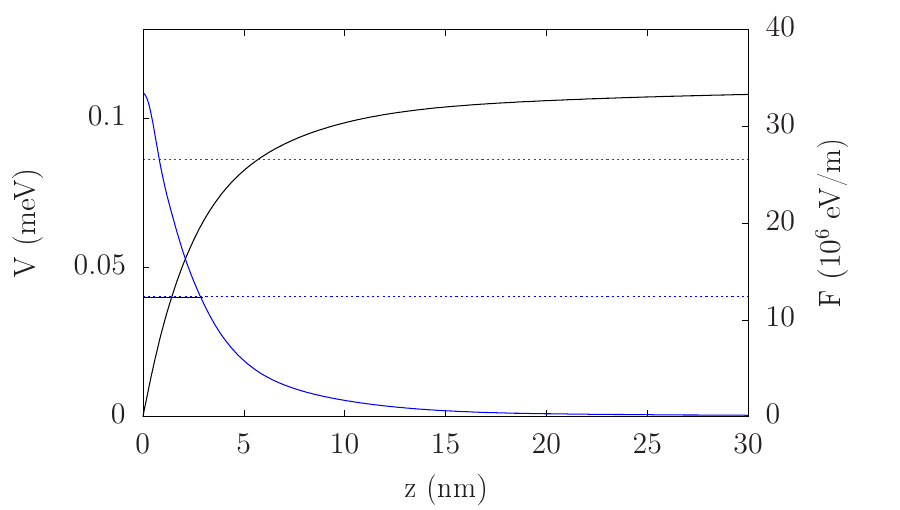}
\end{tabular}
\caption{Self-consistent potential profile at the LAO/STO interface (black line, left axis) together with the corresponding electric field (black curve, right axis). Horizontal dashed lines  mark the energy of the ground state for the $d_{xy}$ (black) and $d_{yz/xz}$ (red) orbitals.} 
\label{afig1}
\end{figure} 
For this purpose, we employed the Schrödinger-Poisson approach, as detailed in Ref.\cite{Scopigno}, utilizing the spatially dependent $\epsilon (z)$ which varies according to the following formula, applicable for low temperatures
\begin{equation}
\epsilon=\epsilon_0+\frac{1}{A+B|F|},
\end{equation}
where $F$ is the electric filed, $A=4.097\times 10^{-5}$, $B=4.907\times 10^{-10}$ m/eV and $\epsilon_0=70$.
In the simulations we took 2DEG electron density at the level $2\times 10^{13}$~cm$^{-2}$ wile the trapped charge profile was assumed in such a way as to achieve an energy difference between the $d_{xy}$ and $d_{xz/yz}$ ground states of approximately 47 meV, as stated in the paper and experimentally measured~\cite{Maniv}.

In Fig.~\ref{afig1} we present the self-consistent potential profile showing that the electric field near the interface where 2DEG is embedded is significant. In this range of about $10$~nm, where wave functions of the ground state for the $d_{xy}$ and $d_{xz/yz}$ orbitals are localized, the dielectric constant changes in the range $(80-200)\epsilon_0$ for the $d_{xy}$ orbital and $(80-500)\epsilon_0$  for the $d_{xz/yz}$ orbital - see Fig.~\ref{afig2}.
\begin{figure}{}
\begin{tabular}{c}
\includegraphics[width=1.00\columnwidth]{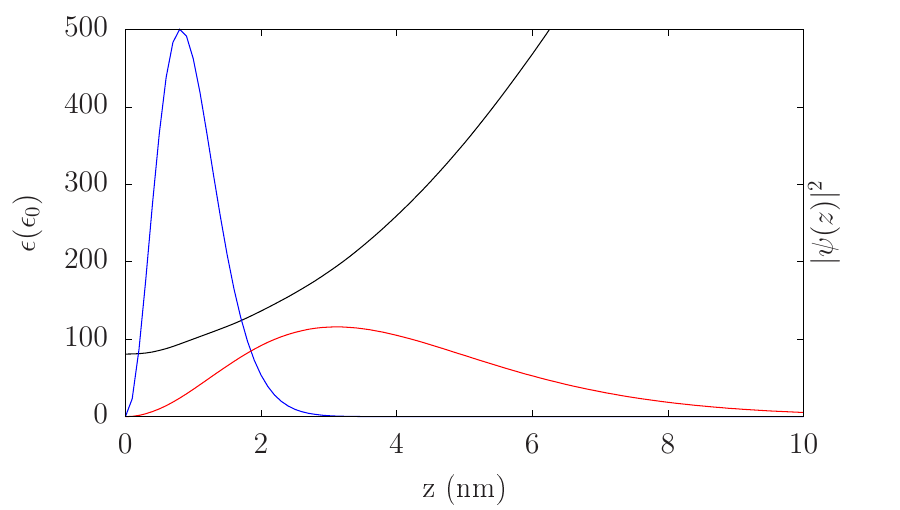}
\end{tabular}
\caption{The dielectric constant as a function of the position at the LAO/STO interface. 
The black and red lines present a square of the wave function for the lowest-energy state of the vertical quantization for $d_{xy}$ (black) and $d_{xz/yz}$ (red) orbitals.} 
\label{afig2}
\end{figure}
As the main contribution to the electronic structure of QD comes from $d_{xy}$ band, due to the shift of this band by 47 meV with respect to $d_{xz/yz}$, we assumed the dielectric constant equal to $100\epsilon_0$ as the average value of $\epsilon$ determined for this band.  Furthermore, we confirmed that varying $\epsilon$ from 100$\epsilon_0$ to 300$\epsilon_0$ does not significantly affect our results.
}

\bibliography{refs.bib}

\end{document}